\documentclass[%
 reprint,
 amsmath,amssymb,
 aps,
]{revtex4-1}

\usepackage{graphicx}
\usepackage{dcolumn}
\usepackage{bm}

\usepackage{algorithm}
\usepackage{algorithmic}
\usepackage[caption=false,labelformat=simple]{subfig}

\DeclareMathOperator*{\argmax}{arg\,max}

\begin{document} 

\preprint{APS/123-QED}

\title{Inequity Aversion and the Evolution of Cooperation}

\author{Asrar Ahmed}
\email{asrar.ahmed@research.iiit.ac.in}
\author{Kamalakar Karlapalem}%
 \email{kamal@iiit.ac.in}
\affiliation{%
International Institute of Information Technology-Hyderabad\\
Centre for Data Engineering \\
Hyderabad, India\\
}%

\begin{abstract}

Evolution of cooperation is a widely studied problem in biology, social science, economics and artificial intelligence. 
Most of the existing approaches that explain cooperation rely on some notion of direct or indirect reciprocity.
These reciprocity based models assume agents recognize their partner and know their previous interactions, which
requires advanced cognitive abilities. 
In this paper we are interested in developing a model that produces cooperation without requiring any explicit memory of previous game plays.
Our model is based on the notion of \emph{inequity aversion}, a concept introduced within behavioral economics, whereby individuals 
care about payoff equality in outcomes.
Here we explore the effect of using income inequality to guide partner selection and interaction. 
We study our model by considering both the well-mixed and the spatially structured population and present the conditions under 
which cooperation becomes dominant. Our results support the hypothesis that inequity aversion promotes cooperative relationship among non kin.

\end{abstract}
\pacs{Valid PACS appear here}
\maketitle

\section{Introduction}
The evolution of cooperative behavior among seemingly competing organisms presents a challenge to the theory of natural selection \cite{axelrod:1984}.
An action that benefits others and comes at a cost to oneself should eventually disappear. On the contrary, there are numerous 
biological and social settings in animal and human societies where cooperation is ubiquitous \cite{dugatkin:1997,sigmund:2006,nowakbook:2006}. 
Evolutionary game theory (EGT) \cite{maynardsmith:1973,maynardsmith:1982} provides a simple framework to study this puzzle. In each generation 
agents play a finite number of games and receive reward according to a 
specified payoff matrix, usually the \emph{prisoner's dilemma} game \cite{rc65}. At the end of each 
generation, players are reproduced proportional to their relative fitness, subject to mutation. Thus EGT mimics natural selection by increasing the relative 
abundance of better performing (in terms of accumulated reward) individuals.

The prisoner's dilemma (PD)   is one of the most widely used games to study the  evolution of cooperation \cite{doebeli_2005,Fogel:1995, boyd:1989, Farrell_Ware:1989}. 
PD is a simultaneous two player game where each player decides whether to  cooperate (C) or defect (D).
Mutual cooperation gives higher payoff than mutual defection. But if one player cooperates the other player is better off 
defecting. Thus each player has an incentive to ``free ride" at a cost to the cooperative player.
In the absence of any special mechanism defecting players would outperform cooperative players and (by natural selection) become dominant \cite{nowak_fiverules}.
The game serves as a metaphor for various real-world settings where there is a conflict between individual and group interest \cite{Hardin_1968_Commons}.

Although many solutions have been put forth to resolve this puzzle, most of these models rely on some notion of direct or indirect reciprocity.
In direct reciprocity \cite{trivers:1971, NowakSigmund1993} agents interact repeatedly, and after each interaction they update their response based 
on how cooperative (kind) or non cooperative (unkind)  the other agent was.
Indirect reciprocity \cite{nowak_id:1998, leimar:2001,nowak_nature:2005} relies on the notion that individuals are kind to those 
who are kind to others. Both models assume agents recognize their partner and know their previous interactions.
In large dynamic settings such an assumption would require advanced cognitive abilities.

Approaches that do not require reciprocity include kin selection and group selection models. In kin selection \cite{hamilton:1964,taylor:1992} agents 
are favorably biased towards their genetic relatives. It requires agents to differentiate between kin and non-kin agents and does not  
explain cooperative outcomes observed in unrelated individuals.
Group selection \cite{traulsen_nowak:2006a,wilson:1975} addresses these issues by proposing that selection works not only 
on individuals but also on groups.
Agents form groups and cooperate with other agents in the same group.

An interesting addition to the above reciprocity-free models of cooperation is the \emph{tag-mediated} partner 
selection \cite{riolo_axelrod:2001,riolo:1997}. Tags are simple observable traits or cultural artifacts which agents  use 
to favorably bias their interactions
with agents having similar tags. Experimental results \cite{riolo:1997} show that a tag based model can produce 
stable cooperation even in single round prisoner's dilemma game.

In this paper, we are interested in developing a model of cooperation that does not require an agent's type or its past interaction to be known explicitly.
Our model is based on the notion that individuals are averse to income inequalities. This 
concept has been formalized and widely studied within behavioral economics as \emph{inequity aversion} \cite{diffavers}.
It was introduced  to account for outcome anomalies between experimental results and theoretical models. 
In the inequity aversion model, as proposed by Fehr and Schmidt \cite{diffavers}, individuals enforce income equality
by forgoing monetary payoff. If the material payoffs of players $i,\ j$ are $x_i,\ x_j$, respectively, the \emph{experienced utility} (which players maximize)
of player $i$ is given by
\[ U_i(x_i,x_j) = x_i - k_1 \cdot max(x_j-x_i,0) - k_2 \cdot max(x_i-x_j,0) \]
where $k_1<k_2$ and $0 \leq k_2 \leq 1$ are inequity aversion sensitivity parameters.  It follows from the above equation that the experienced utility 
is maximized when inequity is zero. 
The condition $k_1 < k_2$ implies that the decrease in a player's utility is higher when it is behind (in terms of payoff) than when it is ahead.
Fehr and Schmidt \cite{diffavers} show that the above model can account for cooperation observed in various human interactions.

Here we study the effect of using the social paradigm of inequity aversion as a criterion for partner selection. Instead of
defining experienced utility we simply allow agents to use accumulated payoff to select their game partner.
Thus inequity aversion in our context implies that agents avoid interacting with other agents whose accumulated payoff
is higher or lower than their own payoff. 
Our experimental results  show that cooperation can emerge even if individuals receive only rudimentary environmental signals about others' well-being 
(and not their type or specific behavior). It supports the hypotheses that inequity aversion promotes cooperation among non kin \cite{evolve_ia}. 
We also observe a strong correlation between cooperation and inequity aversion that indicates a possible coevolution of these two behaviors \cite{evolve_ia}.

The rest of the paper is organized as follows. We  first present the model and define the partner selection bias as a function 
of income inequality. We then evaluate the model by considering both the well-mixed and the spatially structured population and highlight the 
evolutionary dynamics of cooperation and inequity aversion. We also study the effects of various parameters on the fraction of cooperative players in the environment 
and then conclude. 

\section{Model}
In this section we first introduce the general model where agents interact in a well-mixed population. 
We later present the model with spatial constraints. 




\emph{Agents.}
The model consists of a fixed number of agents $N$.  Each agent is either the cooperative or defecting type.
The cooperative type always plays cooperate, and the defecting type always plays defect. 
Agents have an associated parameter $\lambda$ which gives a measure of how sensitive or tolerant  they are to payoff inequality.
A player's type and sensitivity parameter are subjected to evolutionary changes (see evolutionary step below). 
Players also have an accumulated reward which is initialized to zero at the start of each generation and updated according to the payoff received 
in each game.

\emph{Payoff matrix.}
The payoff received by each player is specified by the standard prisoner's dilemma game: 
\[
\begin{pmatrix}
b-c & -c\\
b& 0
\end{pmatrix}.\]

Cooperative players provide benefit $b$ to their opponent and incur a cost $c$ ($b>c>0$). Defecting players neither provide benefit nor incur any cost.
Given the payoff structure, the dominant strategy for each player, irrespective of what the other player does, is to defect.
Thus players end up with a payoff of zero instead of the mutually beneficial payoff $b\ -\ c$. This characterizes the dilemma between 
individual interest and group well-being. Without loss of generality, we use the following normalized payoff matrix: 

\[
\begin{pmatrix}
1 & 0\\
1+c/b & c/b
\end{pmatrix}.\]

This allows us to study the game as a function of a single cost-to-benefit parameter, $0<c/b < 1$ \cite{fu_nowak2010,langer:2008}. 

\begin{figure}[t]
\begin{center}
    \includegraphics[width=0.35\textwidth,height=1.6in]{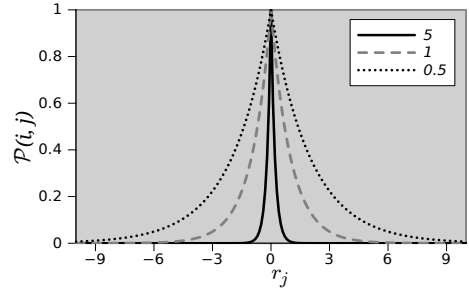}
  \caption{$ {\cal P}(i,j)$ for different values of $r_j$ and $\lambda_i=\{ 5,1,0.5\}$ when $r_i=0$ and $\lambda_j=0$.}
\label{fig:probinter}
\end{center}
\end{figure}

\begin{figure*}[t]
\begin{minipage}[b]{0.74\linewidth}
\begin{minipage}[b]{1.0\linewidth}
\centering
\subfloat[]{\label{fig:coop_across_gen}\includegraphics[width=1.0\textwidth,height=0.75in]{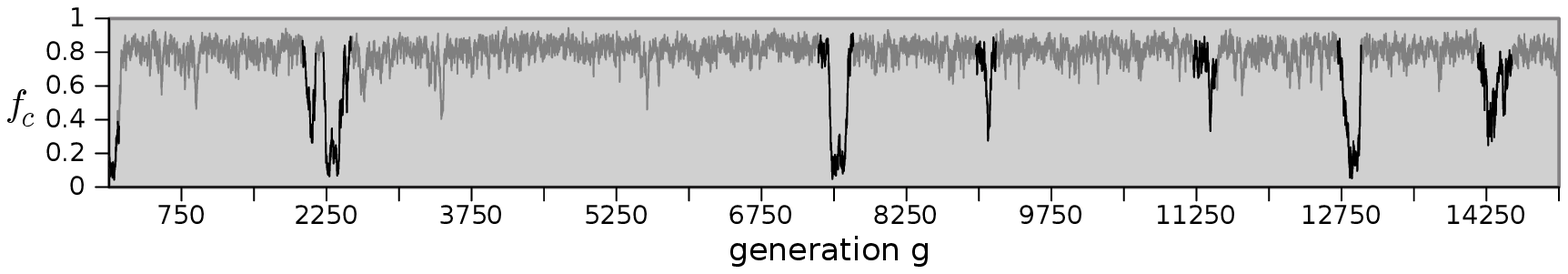}}
\end{minipage}
\begin{minipage}[b]{1.0\linewidth}
\centering
\subfloat[]{\label{fig:lambda_across_gen}\includegraphics[width=1.0\textwidth,height=0.75in]{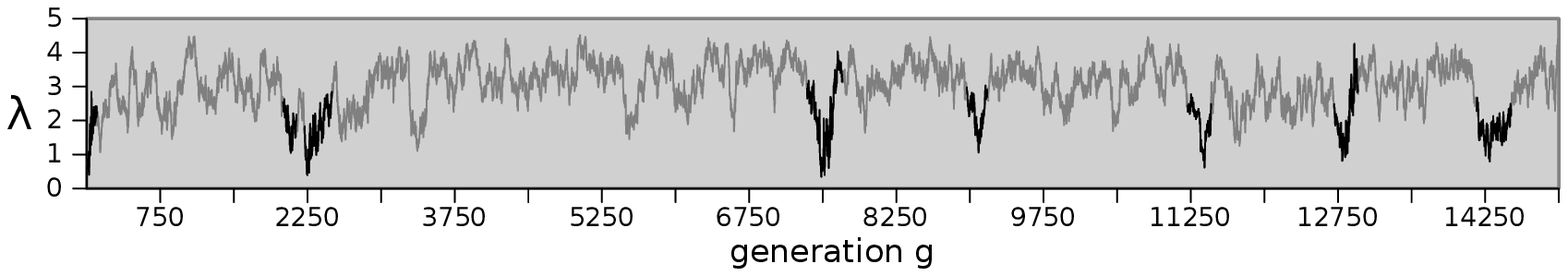}}
\end{minipage}
\end{minipage}
\hspace{0.01cm}
\begin{minipage}[b]{0.24\linewidth}
\centering
\subfloat[]{\label{fig:coop_lambda}\includegraphics[width=1.0\textwidth,height=1.75in]{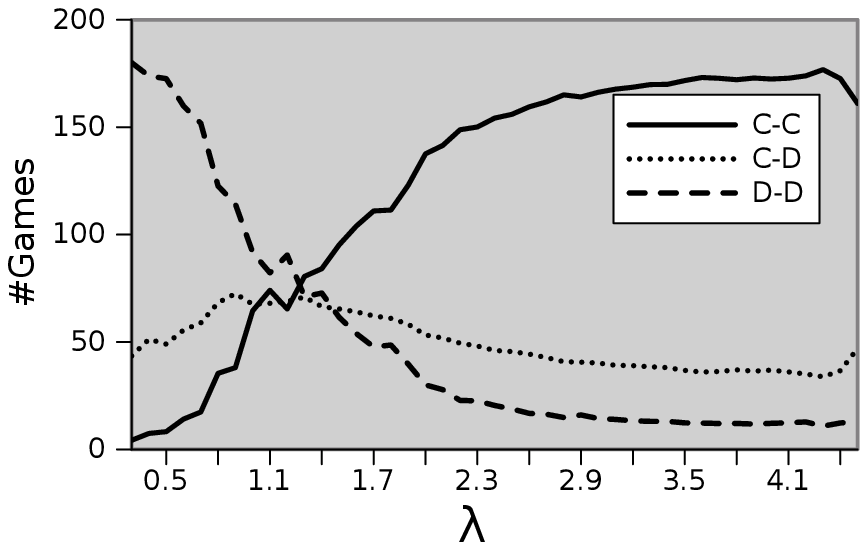}}
\end{minipage}
\caption{(a) The fraction of cooperative players $f_c$ across generations for a typical run for $N=250,\ N_s=8$, and $c/b=0.45$. (b) The corresponding 
average $\lambda$ value of cooperative players. (c) The corresponding number of games with C-C, C-D, and D-D interactions vs $\lambda$.}
\end{figure*}

\emph{Partner selection and interaction.}
In each generation, agents are selected sequentially (random order) and play a single round of the prisoner's dilemma game.
Players use their accumulated payoff and sensitivity parameter to determine their game partner. 
The search space variable $N_s \in N$ gives the subset of agents from which players choose their partner.
Denote  $r_i,\ r_j$ as the accumulated payoff and  $\lambda_i,\lambda_j$ as the sensitivity parameters of players $i,\ j$, respectively.
The probability that player $i$ would accept player $j$ is given by

\[ \phi(i,j) = e^{-\lambda_i |r_i-r_j|} .\]

Given the search space $N_s$, player $i$ selects player $j$ such that the above probability is maximized, $j= \argmax_{j \in N_s} \phi(i,j)$.
We similarly define $\phi(j,i)$ as the probability that player $j$ would accept player $i$ as its partner. 
The probability of interaction between players $i$ and $j$ is given by
\begin{eqnarray}
{\cal P}(i,j) &  = &\phi(i,j)  \phi(j,i) \nonumber \\
	& = & e^{-(\lambda_i+\lambda_j)|r_i-r_j|}.
\label{eqn:probi}
\end{eqnarray}

The interaction is bilateral or with mutual consent since the probability of interaction depends on both players' sensitivity parameters.
We set the range of the sensitivity parameter $\lambda \in [0,5]$. If $\lambda_i=\lambda_j=0$, players are indifferent to inequity, and 
as $\lambda$ increases, they become increasingly inequity averse. 
The upper limit of $5$ ensures that the probability ${\cal P}(i,j)$ is close to zero, even for small payoff differences.
To illustrate, Figure.~\ref{fig:probinter} gives the probability of interaction between players $i$ and $j$  for different values of $\lambda_i$ when 
$i$'s accumulated reward $r_i=0$ and $\lambda_j=0$. For $r_j=1$ and $\lambda_i=0.5$, the probability of interaction ${\cal P}(i,j)=0.6$
which decreases to ${\cal P}(i,j)=0.006$ for $\lambda_i=5$.

Once the partner is selected, the PD game is played with the probability given in Equation.~(\ref{eqn:probi}). If both players
are cooperative, their accumulated payoff increases by $1$. If only one of the players is cooperative, 
the defecting player's accumulated payoff increases by $1+c/b$, and the cooperative player's accumulated payoff remains unchanged. 

We note that, unlike tag-mediated models of cooperation \cite{riolo:1997} where an agent's tags remain fixed in a given generation, in our model 
the accumulated payoff serves as a dynamic tag which changes after each interaction.

\emph{Evolutionary step.}
At the end of each generation agents are reproduced using the \emph{binary tournament } procedure \cite{suzuki:2005,goldberg:1991}: 
\begin{list}{}{\labelsep=0.3em \labelwidth=1em \leftmargin=1.5em \parsep -0em \partopsep -0.0em \itemsep 0.2em}
\item [(1)] Two distinct agents are randomly selected for a tournament, and the agent with the higher fitness value is declared the winner  
(we use the accumulated reward as the fitness value).  
\item[(2)] A copy of the winner, called the \emph{offspring}, is added to the new generation and the above procedure is 
repeated  $N$ times.
\end{list}
Additionally, a mutation is applied to each offspring. The mutation value gives the probability with which the offspring's type is randomly reset to either 
the cooperative or defecting type. 
Since the sensitivity parameter is a continuous variable, we apply mutation 
by adding, with probability $\mu$, Gaussian noise with mean $0$ and deviation $1$ to the inherited $\lambda$ value \cite{riolo_axelrod:2001}. 
The accumulated reward of the offspring is set to zero. The algorithm in Table~\ref{algo:a_interact} provides the pseudo code of the model. 

\begin{algorithm}[H]
\caption{Algorithm for agent interaction}
\begin{algorithmic}
\REQUIRE $\mu,N,N_s$.
\WHILE {generation $g \leq g^{max}$ }
\FORALL{agents $i$ $\in$ $N$  }
\STATE {select $N_s \in N$ agents randomly}
\STATE $j= \argmax_{j \in N_s} \phi(i,j)$ 
\STATE{With Probability ${\cal P}(i,j)$, \textsc{Play}$(i,j)$}
\STATE {Update accumulated payoff $r_i,r_j$}
\ENDFOR
\STATE {New Set Of Players $N$ = EvolutionaryStep($N$, $\mu$)}
\ENDWHILE
\end{algorithmic}
\label{algo:a_interact}
\end{algorithm}

\section{Results and Discussion}
We recall that $N$ is the number of agents, $N_s$ is the search space, and 
$c/b$ is the cost-to-benefit ratio. We denote the fraction of cooperative players with $f_c$. C-C denotes the cooperative-cooperative player interaction. 
We similarly define C-D and D-D interactions. 
For all experiments, we set $\mu=0.1$ and initialize $\lambda$ to a uniform value between $[0,5]$. The initial fraction of cooperative players is set to $10\%$.
We report the results by averaging across $20$ runs, with each run consisting of $15000$ generations. Our results
cover the following aspects: (1) the evolutionary dynamics of $f_c$, $\lambda$, and the  correlation between them, 
(2) the change in the fraction of games with C-C, C-D, and D-D interactions as $\lambda$  value changes, 
(3) the effect of number of agents  and search space  on  $f_c$ as the cost-to-benefit ratio $c/b$ increases, and
(4) the effect of spatial constraints on $f_c$.

\begin{figure*}[t]
  \centering
  \subfloat[]{\label{fig:coopN}\includegraphics[width=0.32\textwidth,height=1.78in]{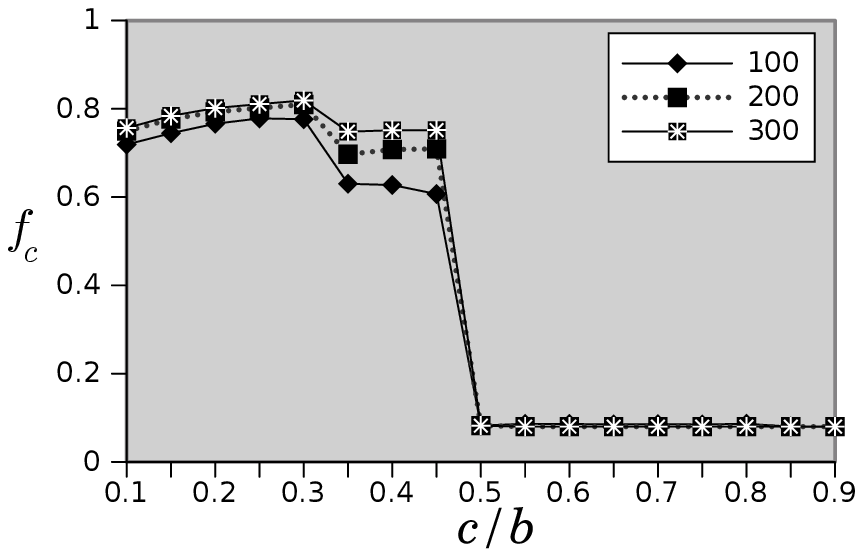}}\hspace{0.5em}
  \subfloat[]{\label{fig:coopNs}\includegraphics[width=0.32\textwidth,height=1.78in]{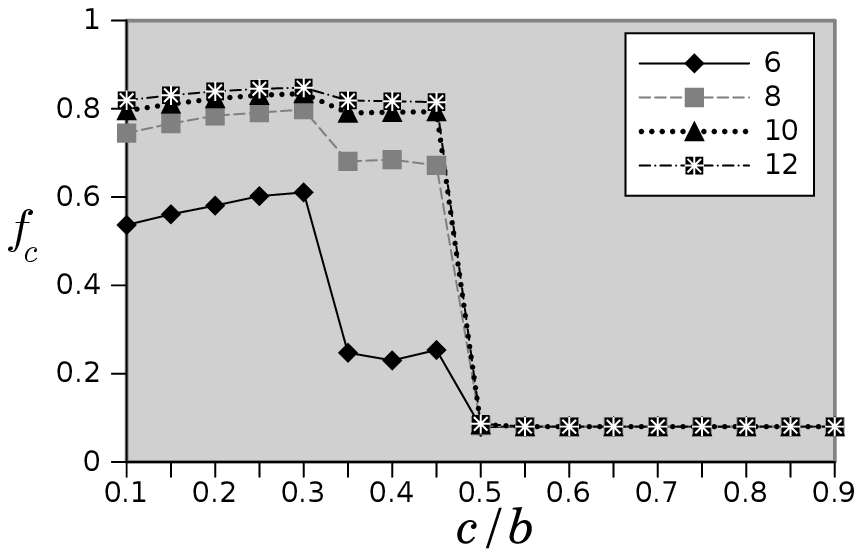}}\hspace{0.5em}
  \subfloat[]{\label{fig:htmap}\includegraphics[width=0.32\textwidth,height=1.81in]{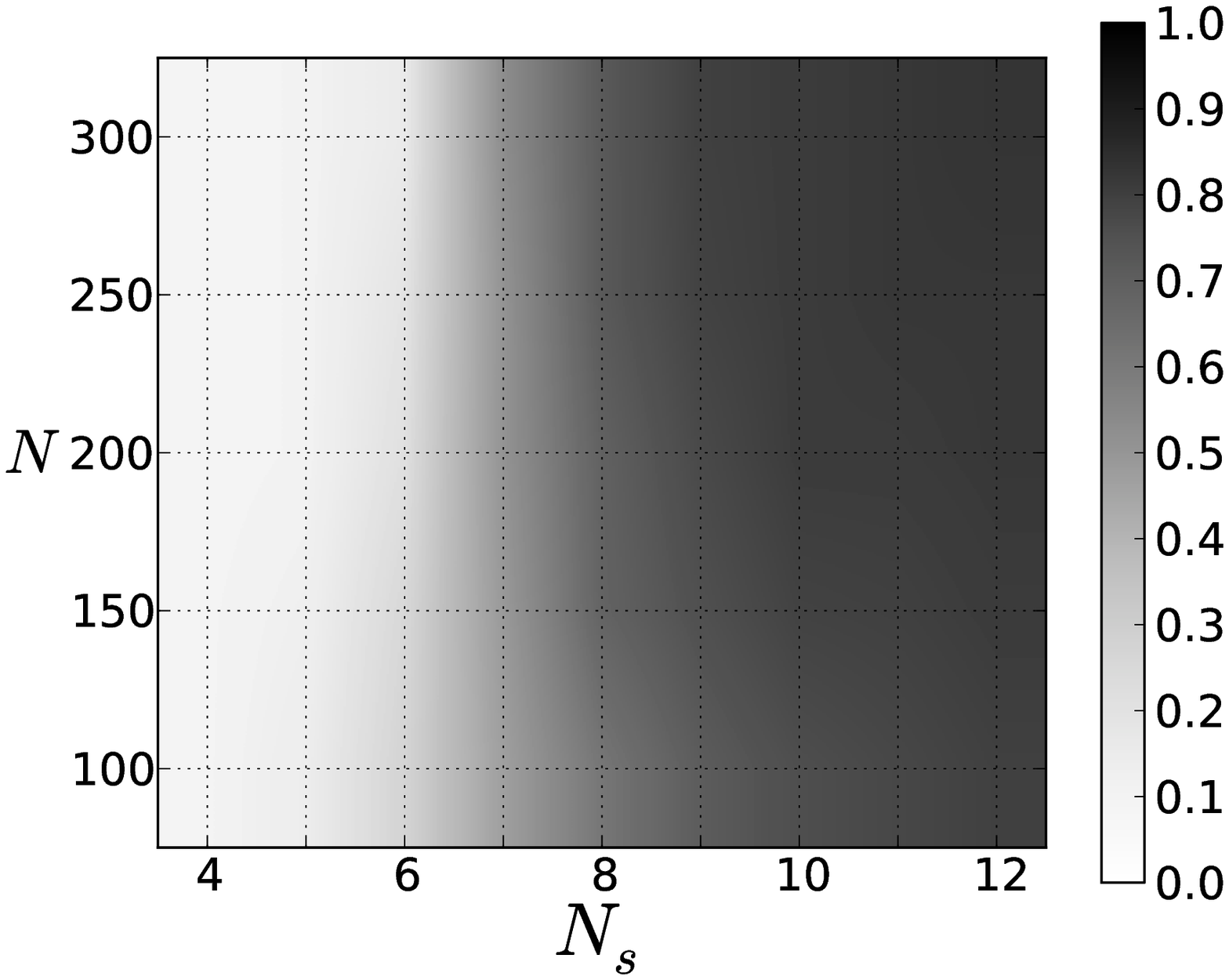}}
  \caption{(a) \emph{f$_c$} vs $c/b$ for $N_s=8$ and $N=\left\{100,200,300 \right\}$. (b)  $f_c$ vs $c/b$ for $N=150$ and $N_s=\left\{6,8,10,12\right\}$. 
	(c) Color map depicting $f_c$ on the $N_s-N$ plane for $c/b=0.4$. }
  \label{fig:params}
\end{figure*}


\subsection{Mixed population}
Figure~\ref{fig:coop_across_gen} shows the fraction of cooperative players $f_c$ across generations for a typical run (with $N=250,\ N_s=8$ 
and $c/b=0.45$). Figure~\ref{fig:lambda_across_gen} gives the corresponding average $\lambda$ value of cooperative players. 
We observe high levels of cooperation $f_c \geq 0.75$ interrupted by brief periods of defection. 
The generations when the fraction of cooperative players falls significantly are highlighted to show the close correlation between $f_c$ and $\lambda$. 
Across all simulations we observe that
an increase or decrease in the fraction of cooperative players  is preceded by an increase or decrease in the average $\lambda$ 
value. For example, as shown in Figures~\ref{fig:coop_across_gen} and ~\ref{fig:lambda_across_gen}, from generation $7100$ to $7500$, the average $\lambda$ value 
decreases gradually from $4.2$ to $0.3$. We see a corresponding decrease in the $f_c$ value from $0.84$ to  $0.08$. 

At the start of each generation, as the accumulated payoff of all players is $0$, the probability of interaction between any two players is $1$, 
irrespective of $\lambda$. So in the initial stages, defecting players have a slight advantage (since the cooperative players would not reject them as a game partner),
and their accumulated payoff increases.
But after the first few games, the cooperative agents involved in C-D interactions 
start forming  temporary clusters sharing a common payoff. If these cooperative players are sufficiently inequity averse, in the subsequent iterations, they 
are more likely to interact within these clusters, and their payoff increases. The defecting players who initially exploit the cooperative players
face a form of ``social exclusion" from these groups. Thus inequity aversion allows cooperative players to seek new partners and form groups
with whom they can share more equitable payoffs. 
As the interactions proceed, cooperative players that are more tolerant (small $\lambda$ value) to income inequality continue to interact 
with defecting agents. They are outperformed by cooperative players with relatively high $\lambda$ value and die out.
Over generations the average $\lambda$ of cooperative players increases, which further reduces the likelihood of C-D interactions
until cooperation becomes dominant.

However, this cooperation due to the temporary clustering effect is not permanent. When cooperation is established, agents have a high $\lambda$ value.
At this stage, the cooperative agents with a relatively smaller $\lambda$  value
have a slight advantage as they are more likely to tolerate inequality and play the game. 
And since cooperative agents are dominant, with high probability these agents interact with other 
cooperative agents, and their accumulated payoff increases. The  environment faces a  slight selection pressure towards higher
tolerance levels (lower $\lambda$ value). As this happens gradually over generations, at some threshold $\lambda$, agents become vulnerable to invasion by defectors,
and the cooperation levels fall sharply. The system remains in this state until by chance, due to mutation, a few cooperative players
with a relatively high $\lambda$ value emerge and reestablish cooperation.

To further validate the correlation between $f_c$ and $\lambda$, Figure~\ref{fig:coop_lambda} shows the number of 
games with C-C, C-D, and D-D interactions as a function of $\lambda$. For a low $\lambda$ value C-D and  D-D 
interactions dominate. As the $\lambda$  value increases, cooperative clusters emerge, and C-C becomes dominant.

The cycles of cooperation and defection are not regular or periodic, and how often the system goes into the defection state and how quickly it recovers
depends on the parameters $N,\ N_s$, and $c/b$. In general we observe that as the cost-to-benefit ratio $c/b$ increases or $N_s$ decreases, it takes longer for 
the system to recover from the defection state.

\begin{figure*}[t]
\begin{minipage}[b]{0.48\linewidth}
\begin{minipage}[b]{1.0\linewidth}
\centering
\subfloat[]{\label{fig:2dcoop_across_gen}\includegraphics[width=1.0\textwidth,height=0.6in]{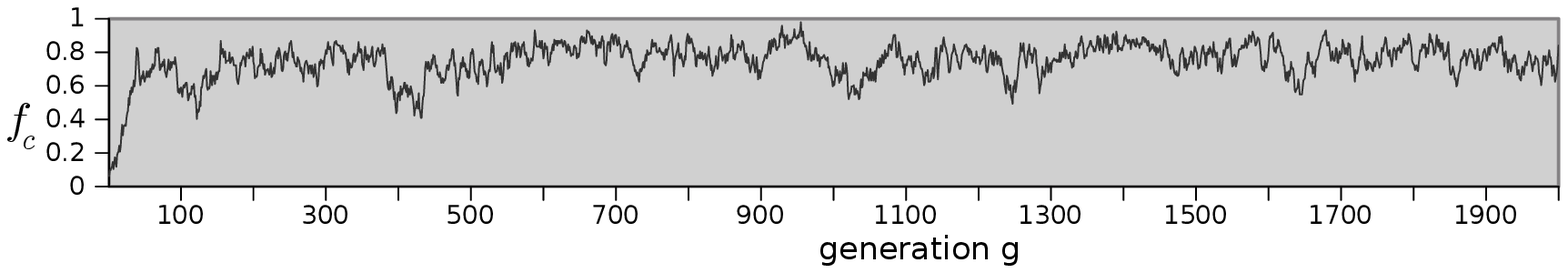}}
\end{minipage}
\begin{minipage}[b]{1.0\linewidth}
\centering
\subfloat[]{\label{fig:2dlambda_across_gen}\includegraphics[width=1.0\textwidth,height=0.6in]{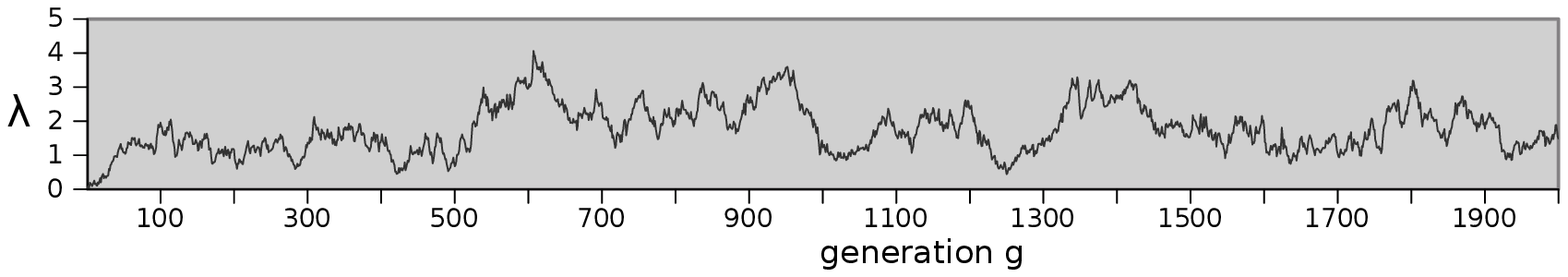}}
\end{minipage}
\end{minipage}
\hspace{0.01cm}
\begin{minipage}[b]{0.24\linewidth}
\centering
\subfloat[]{\label{fig:2dwithin_across}\includegraphics[width=1.0\textwidth,height=1.45in]{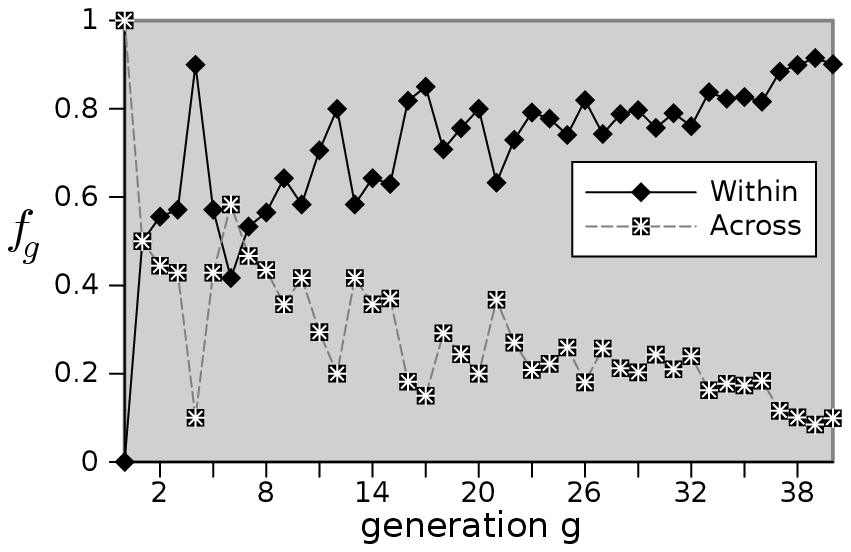}}
\end{minipage}
\hspace{0.01cm}
\begin{minipage}[b]{0.24\linewidth}
\centering
\subfloat[]{\label{fig:2ddiffn}\includegraphics[width=1.0\textwidth,height=1.45in]{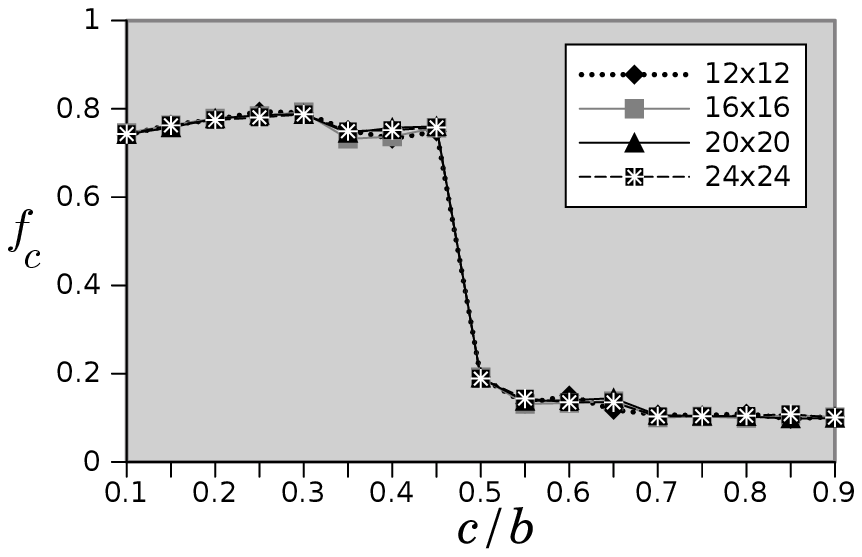}}
\end{minipage}
\caption{(a) The fraction of cooperative players $f_c$ across generations for a typical run with $N=12\times12$ and $c/b=0.45$. (b) The corresponding
average $\lambda$ value of cooperative players. (c) The corresponding fraction of games ($f_g$) within and outside the clusters for the first few generations. 
(d) $f_c$ vs $c/b$ for different grid sizes.}
\end{figure*}

Figures~\ref{fig:coopN} and ~\ref{fig:coopNs} show the fraction of cooperative players vs the cost-to-benefit ratio
for different values of $N$ and $N_s$, respectively.  Across both simulations, as $c/b$ increases, $f_c$ decreases, and for $c/b \geq 0.5$, cooperation disappears. 
With respect to $N$, we observe the 	$f_c$ value decreases marginally for $0.3 \leq c/b \leq 0.5$ as $N$ decreases. 
The change in $f_c$  is considerably higher with respect to change in $N_s$ values. As the search space increases, $f_c$  increases.
The search space affects the probability of selecting an agent from within the temporary clusters that emerge. A higher search space value
translates to a higher probability of interaction within the cluster. We also observe a ``thresholding'' effect; that is, for a fixed change 
in $N_s$ the increase in the $f_c$ 
value is higher for smaller values of $N_s$. 
Figure~\ref{fig:htmap} shows a  color map of $f_c$ for different values of $N_s$ (along the 	$X$ axis) and $N$ (along the $Y$ axis) with $c/b=0.45$. 
For a high ($N_s \geq 12$) or low search space ($N_s \leq 4$), $f_c$ does not change with $N$.
\subsection{Two-dimensional  lattice} We now consider the spatial prisoner's dilemma game \cite{szabo_tiffmode:1998,nowak_robert:1992} where each player occupies 
a cell in a  square lattice. Similar to the well-mixed population model, agents are either the cooperative or defecting type and use 
their accumulated payoff  to select their game partner and interact. 
But due to spatial constraints an agent's search space is restricted to its four neighboring cells. 
We also change the \emph{binary tournament} procedure to reflect an agent's fixed position. 
For each offspring cell to be added to the new generation, two distinct agents are randomly selected from the neighborhood of the cell.
The agent with the higher fitness values is declared the winner and the offspring inherits the winner's type and $\lambda$ value.
Mutation is applied as discussed in the previous model.

Figures~\ref{fig:2dcoop_across_gen} and ~\ref{fig:2dlambda_across_gen} show the evolutionary dynamics of $f_c$ and $\lambda$ 
for a typical run with $N=12\times12$ and $c/b=0.45$ for the first $2000$ generations (similar behavior is seen in  
subsequent generations). Across all simulations, we observe that the cooperation in the spatial prisoner's dilemma
is more ``robust" (i.e., unlike the previous model where cooperation almost disappears before recovering, here we see stable cooperation). 
The increase and decrease in the $f_c$ value are marginal but are  still closely correlated 
with the increase and decrease in the $\lambda$ value.

\begin{figure*}[t]
  \centering
  \subfloat{\label{fig:snap0}\includegraphics[width=0.11\textwidth,height=0.8in]{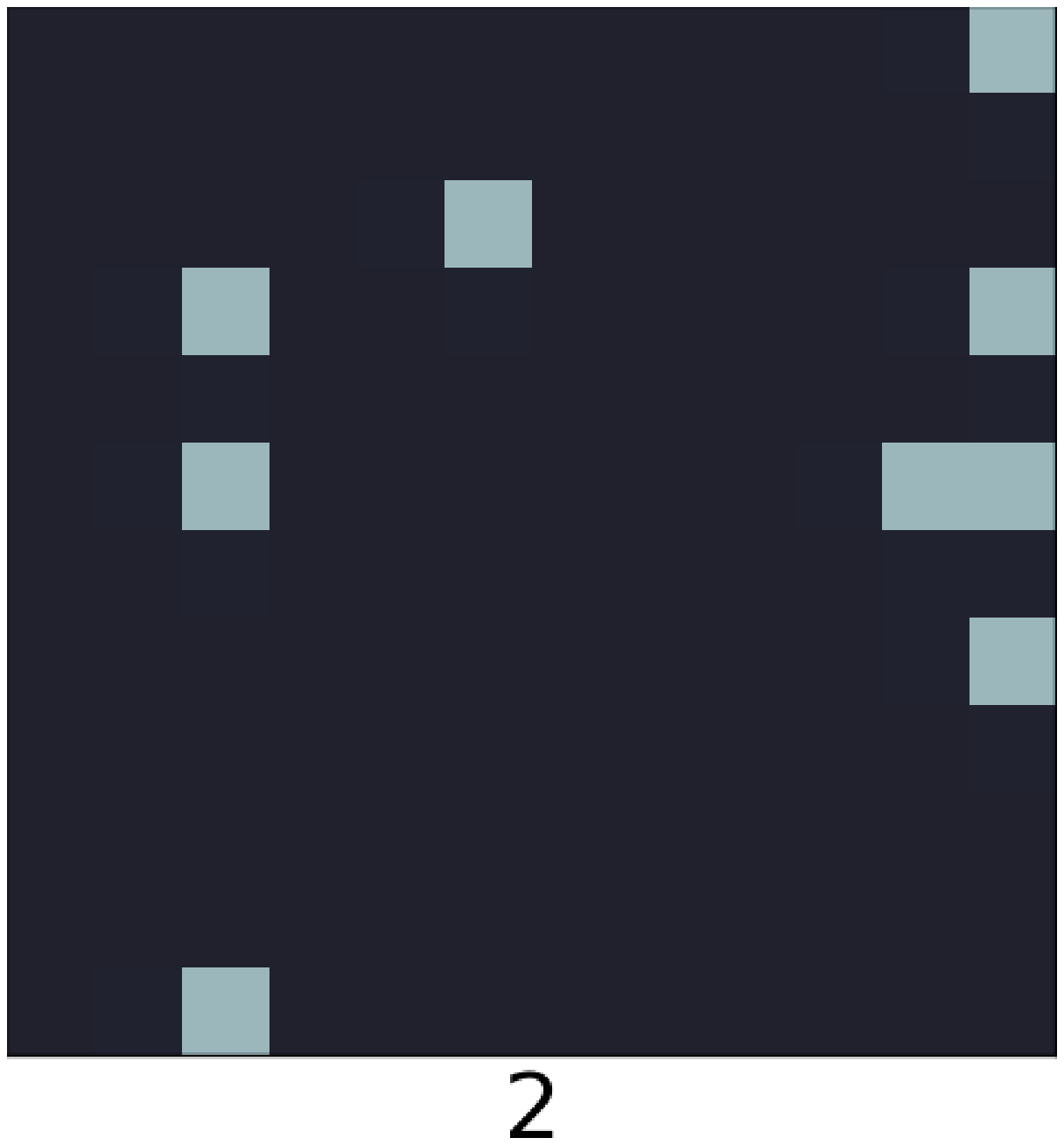}}\hspace{0.4em}
  \subfloat{\label{fig:snap1}\includegraphics[width=0.11\textwidth,height=0.8in]{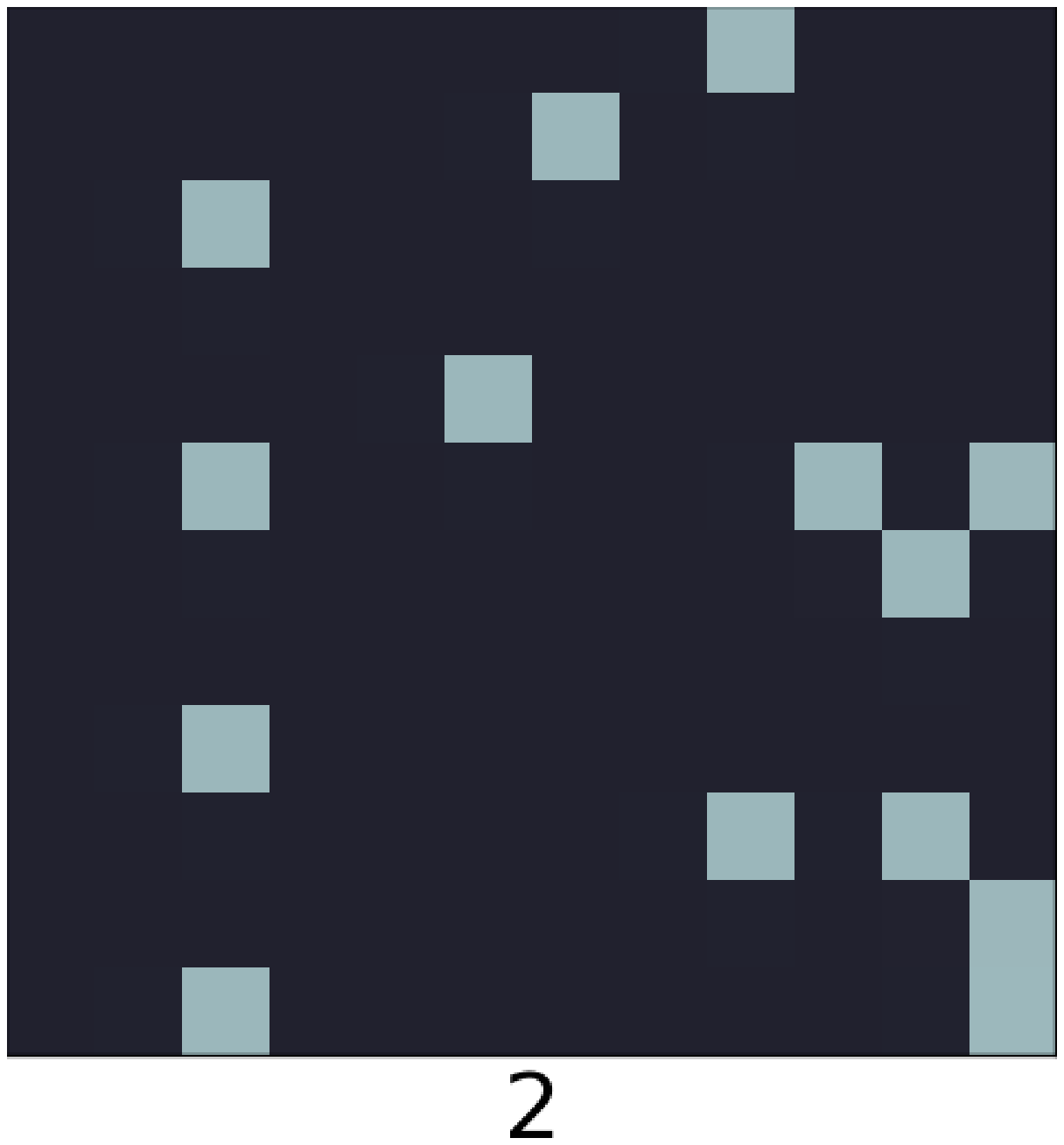}}\hspace{0.4em}
  \subfloat{\label{fig:snap2}\includegraphics[width=0.11\textwidth,height=0.8in]{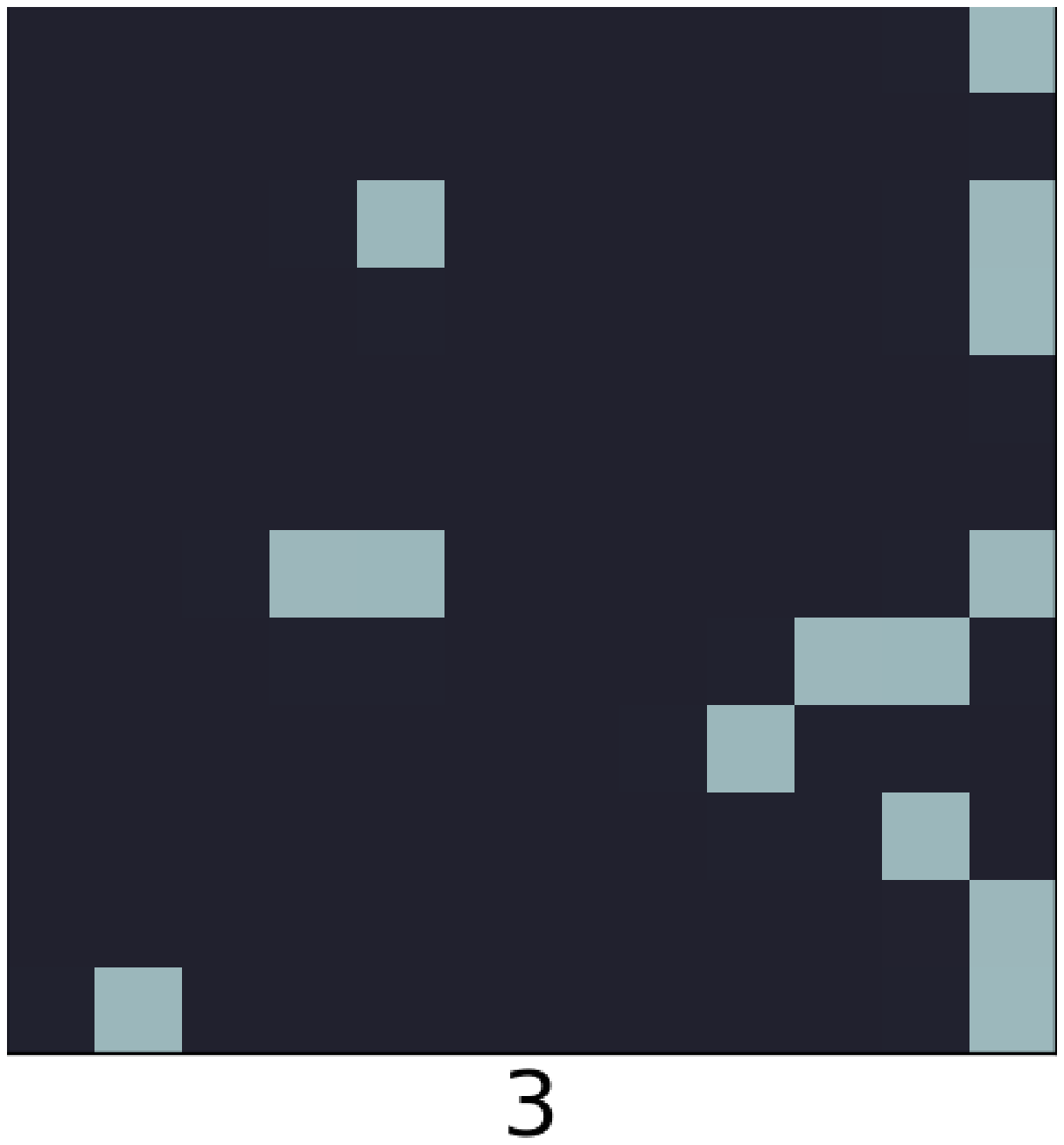}}\hspace{0.4em}
  \subfloat{\label{fig:snap3}\includegraphics[width=0.11\textwidth,height=0.8in]{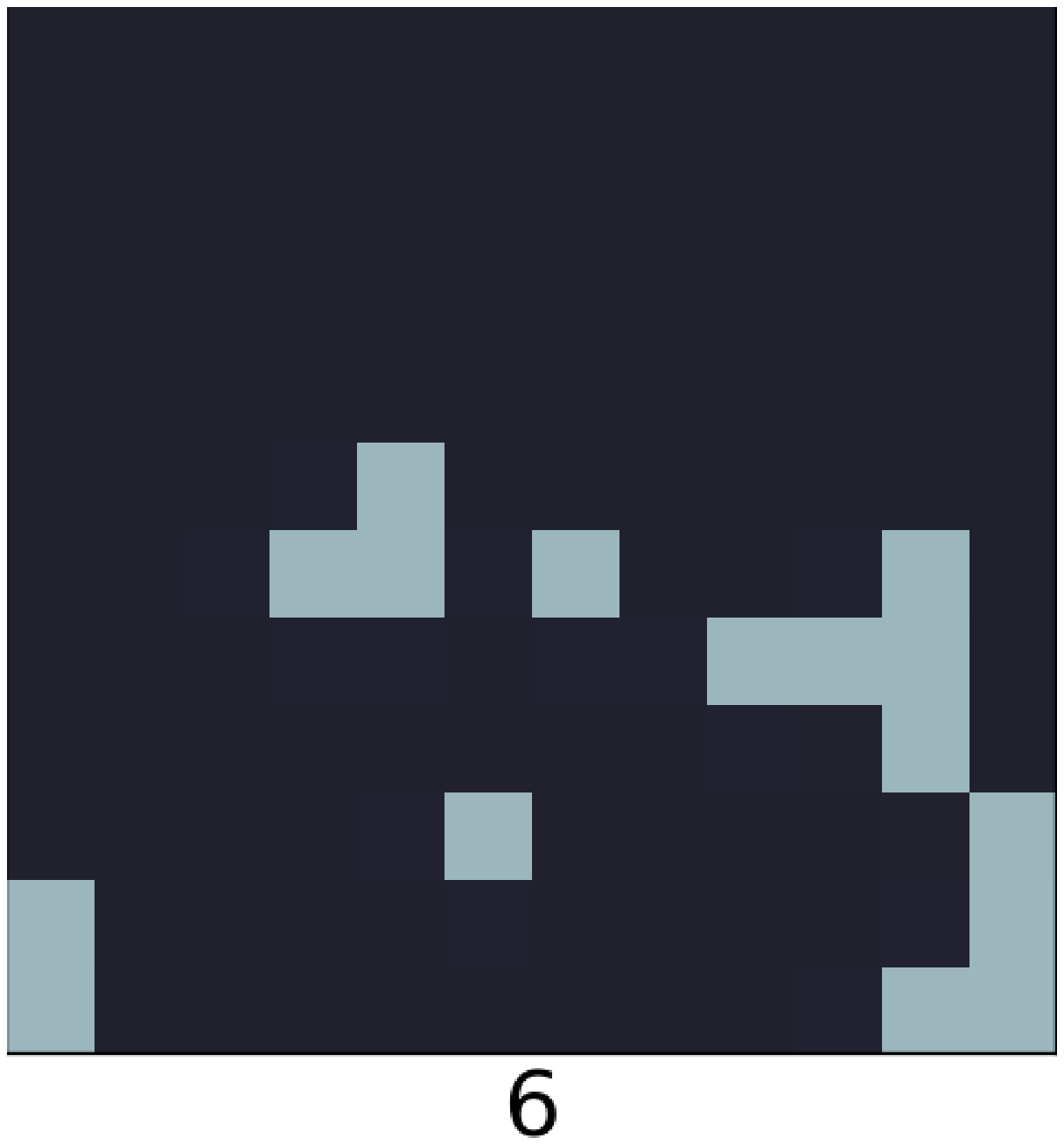}}\hspace{0.4em}
  \subfloat{\label{fig:snap4}\includegraphics[width=0.11\textwidth,height=0.8in]{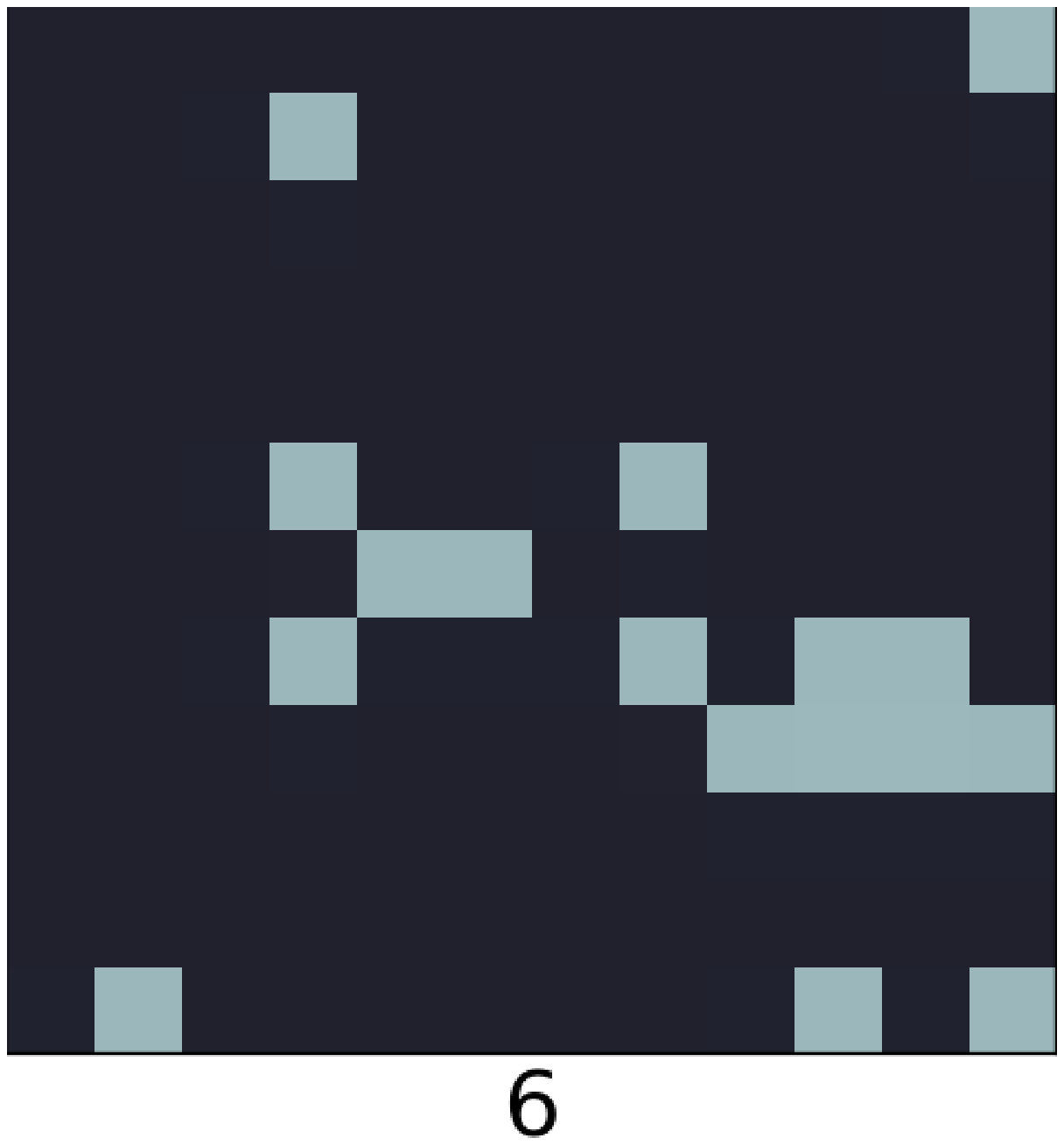}}\hspace{0.4em}
  \subfloat{\label{fig:snap5}\includegraphics[width=0.11\textwidth,height=0.8in]{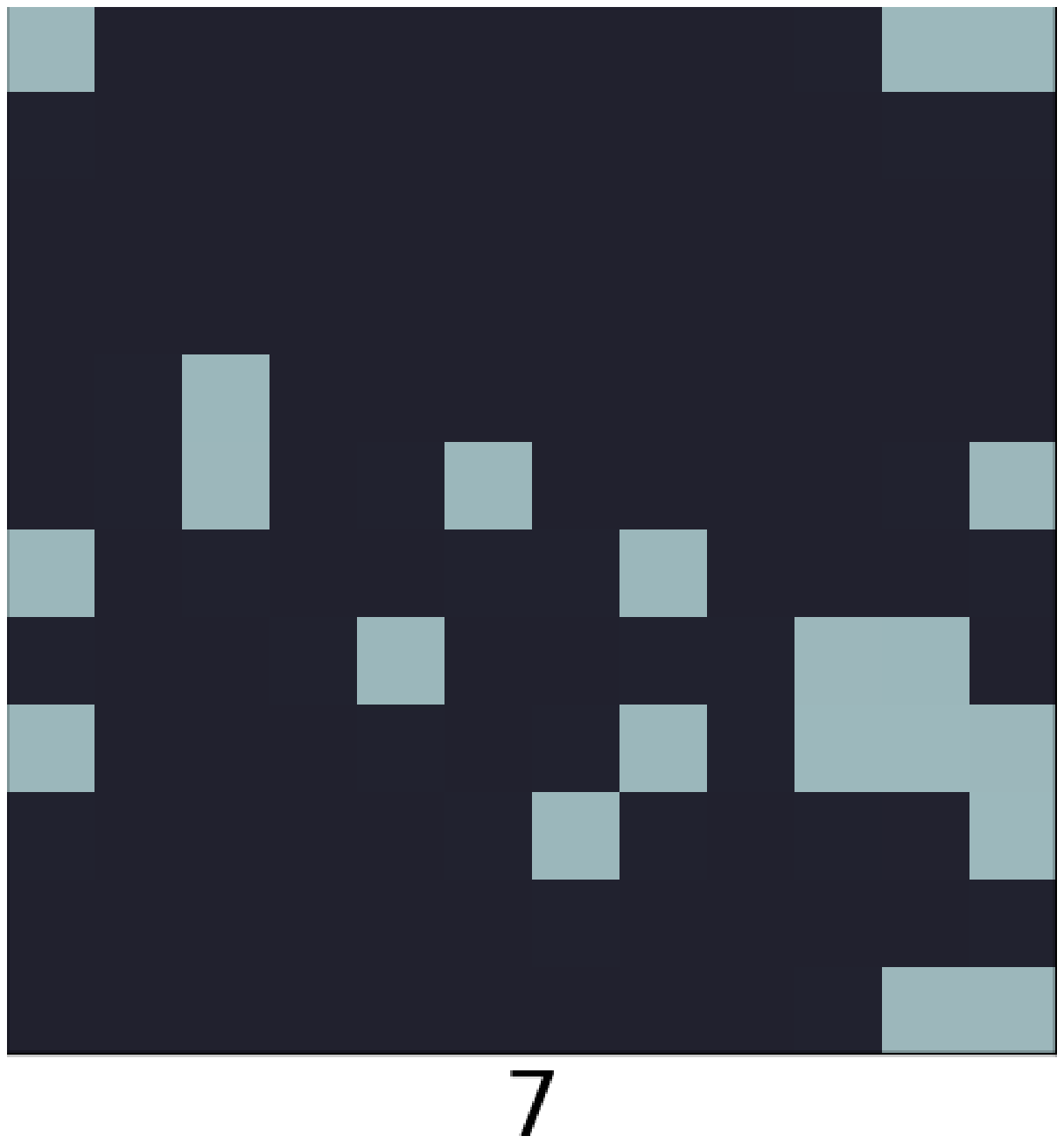}}\hspace{0.4em}
  \subfloat{\label{fig:snap6}\includegraphics[width=0.11\textwidth,height=0.8in]{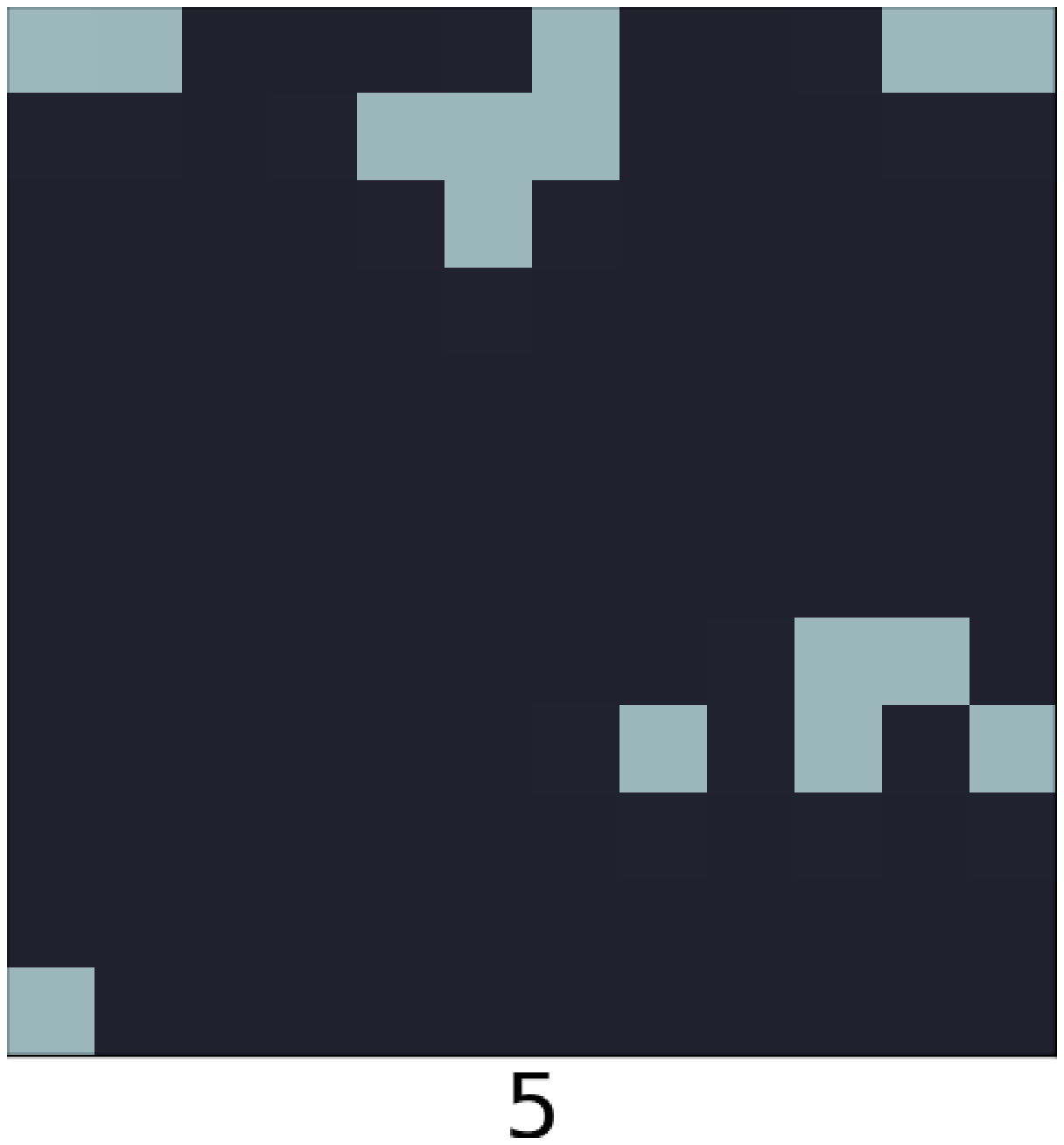}}\hspace{0.4em}
  \subfloat{\label{fig:snap7}\includegraphics[width=0.11\textwidth,height=0.8in]{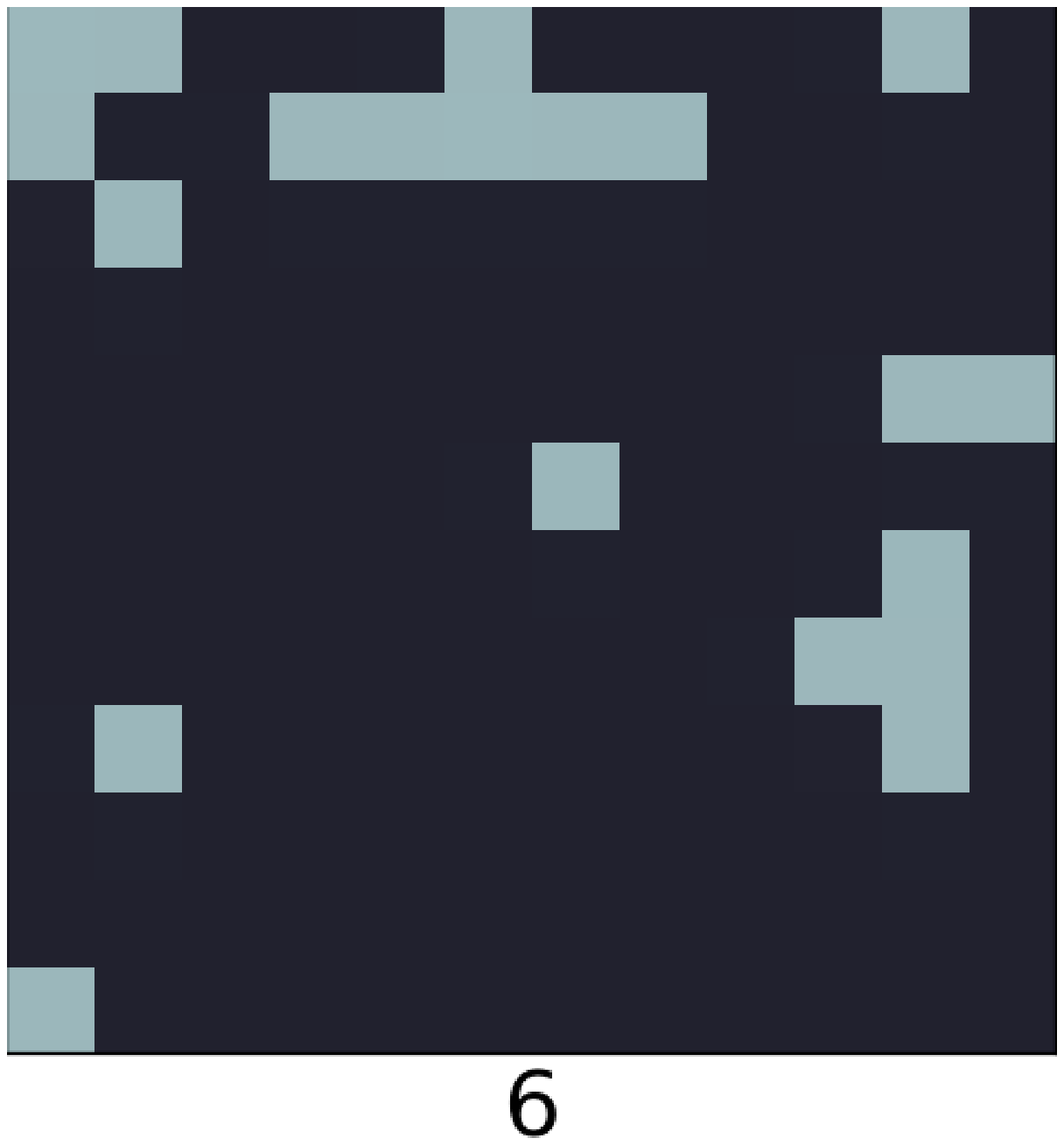}}\hspace{0.4em}\\

  \subfloat{\label{fig:snap8}\includegraphics[width=0.11\textwidth,height=0.8in]{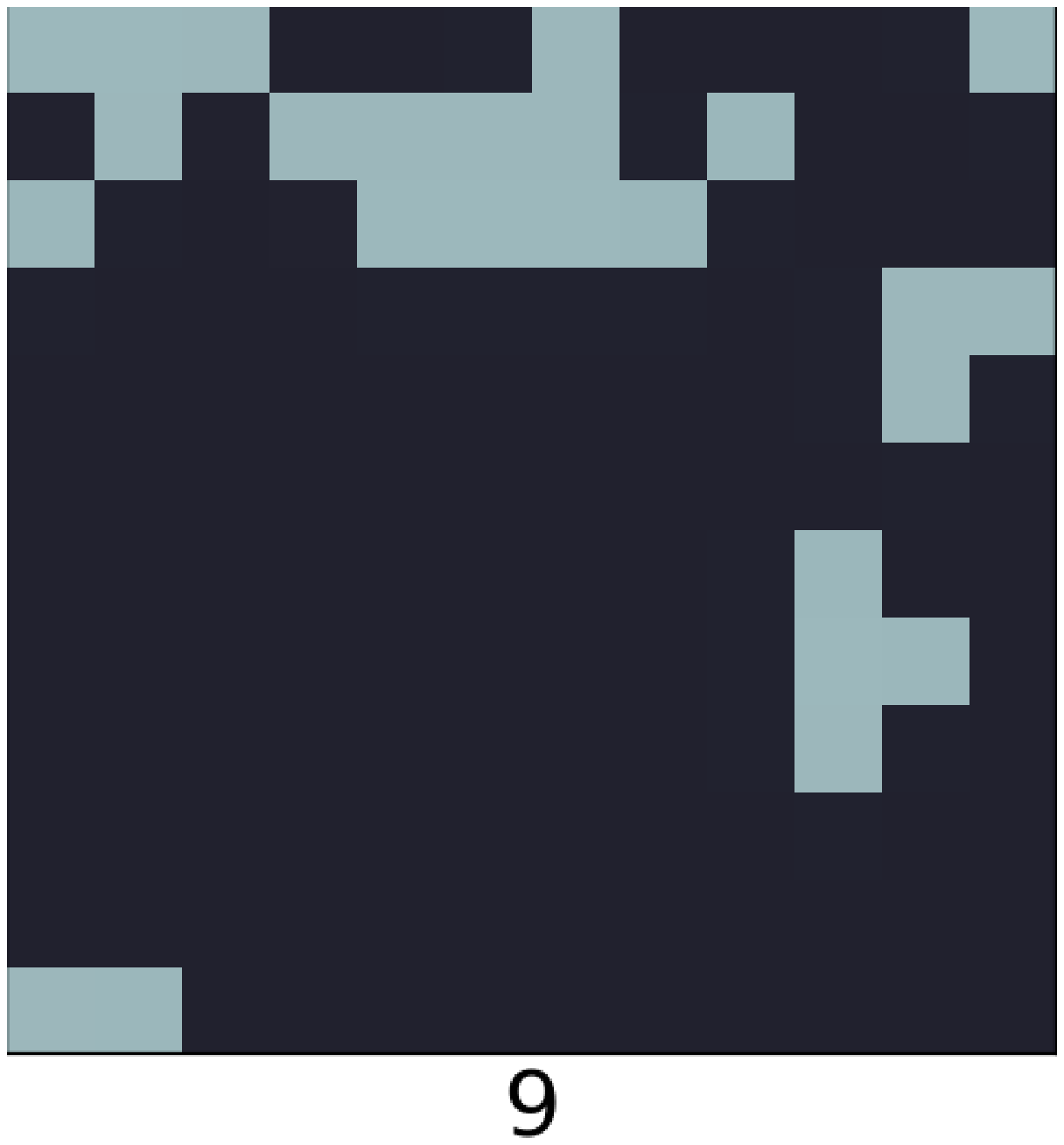}}\hspace{0.4em}
  \subfloat{\label{fig:snap9}\includegraphics[width=0.11\textwidth,height=0.8in]{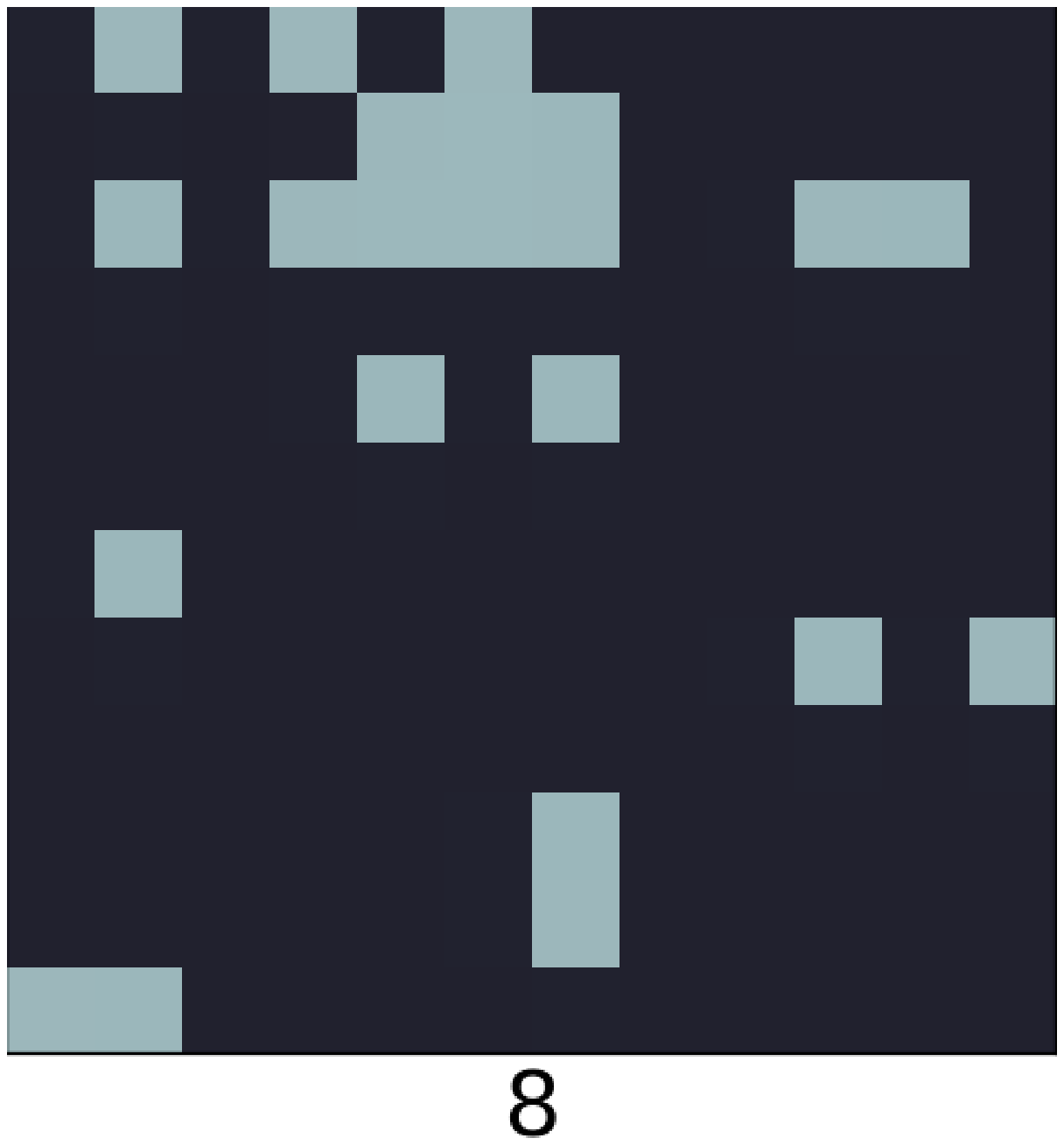}}\hspace{0.4em}
  \subfloat{\label{fig:snap10}\includegraphics[width=0.11\textwidth,height=0.8in]{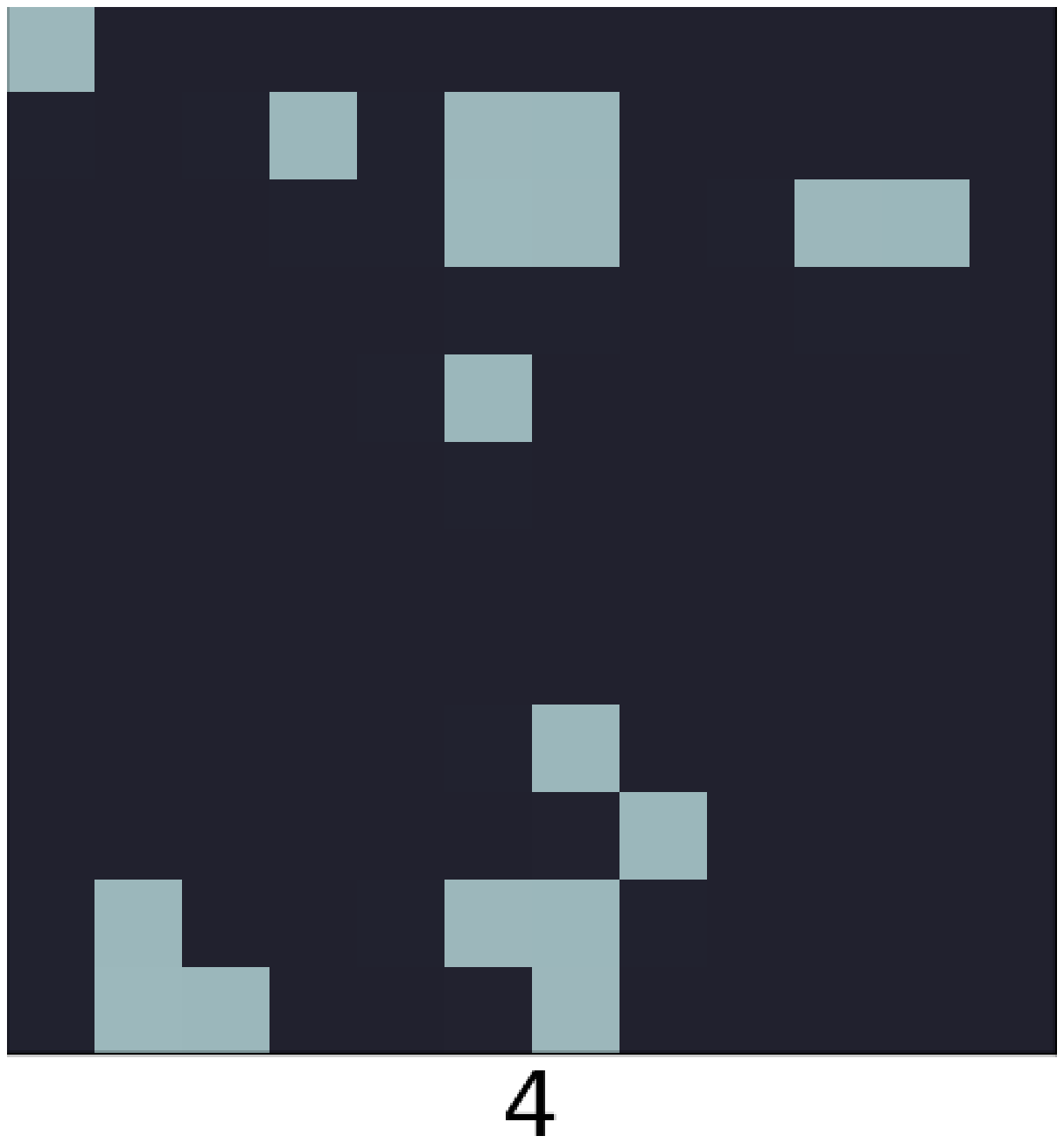}}\hspace{0.4em}
  \subfloat{\label{fig:snap11}\includegraphics[width=0.11\textwidth,height=0.8in]{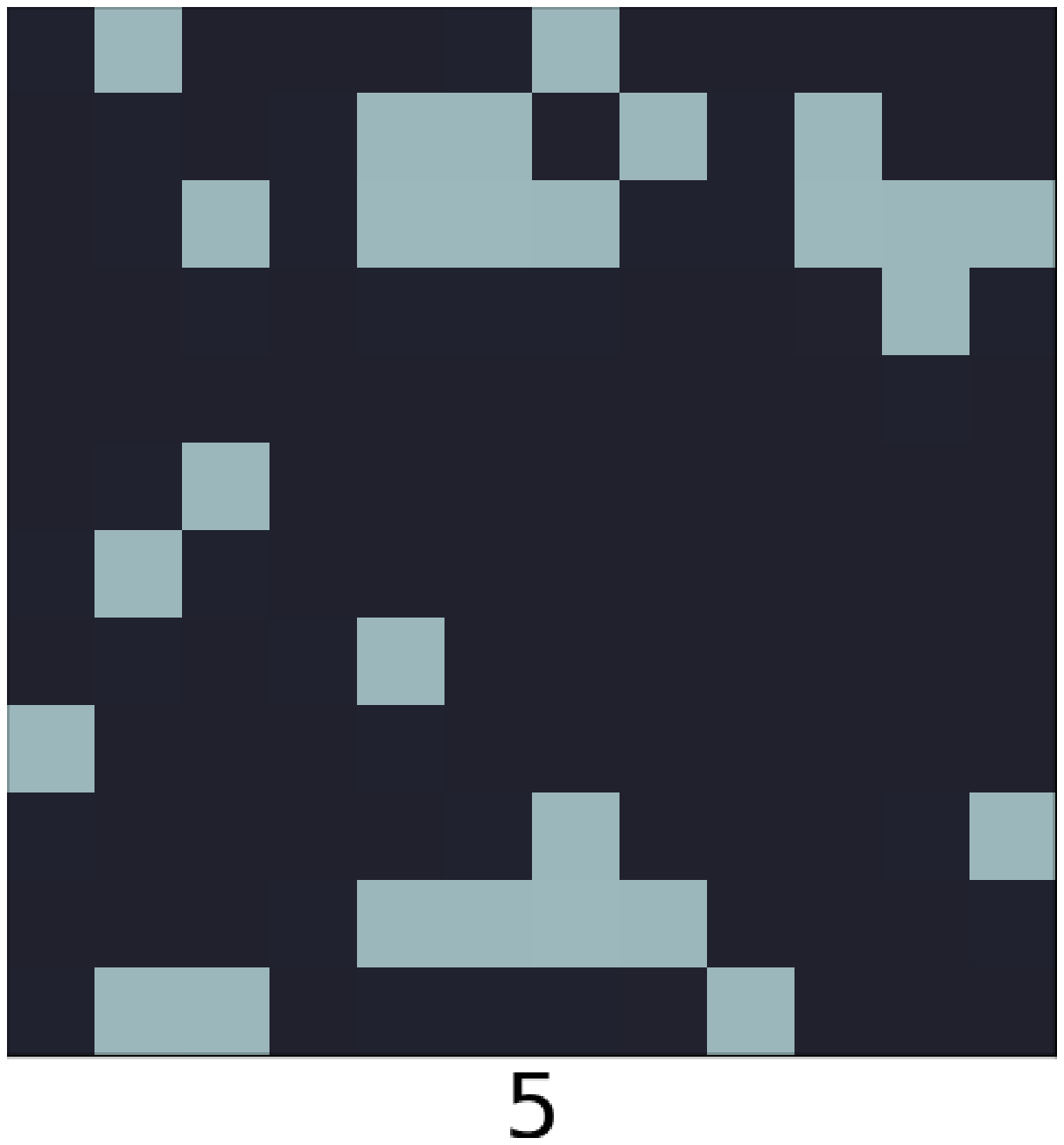}}\hspace{0.4em}
  \subfloat{\label{fig:snap12}\includegraphics[width=0.11\textwidth,height=0.8in]{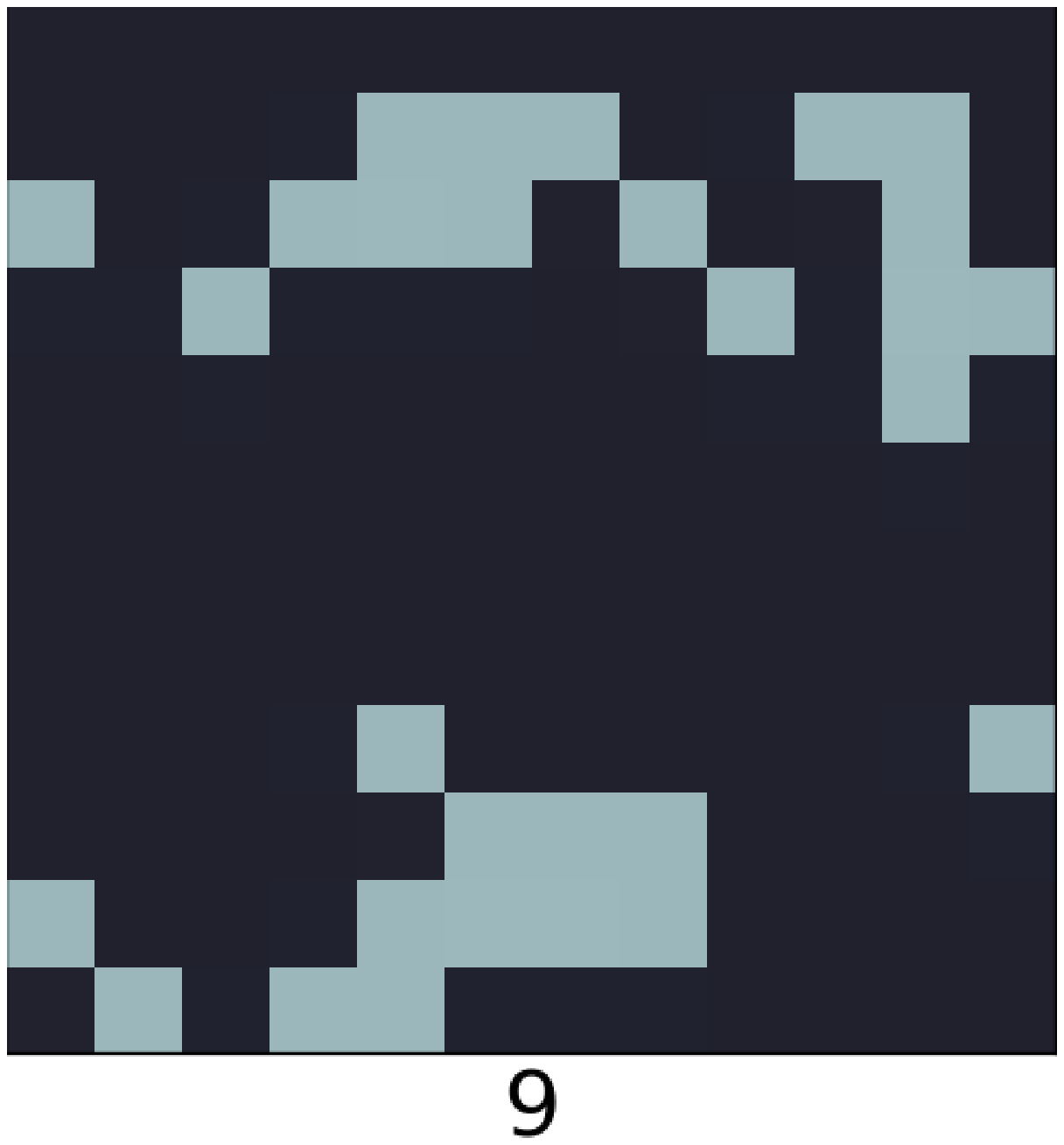}}\hspace{0.4em}
  \subfloat{\label{fig:snap13}\includegraphics[width=0.11\textwidth,height=0.8in]{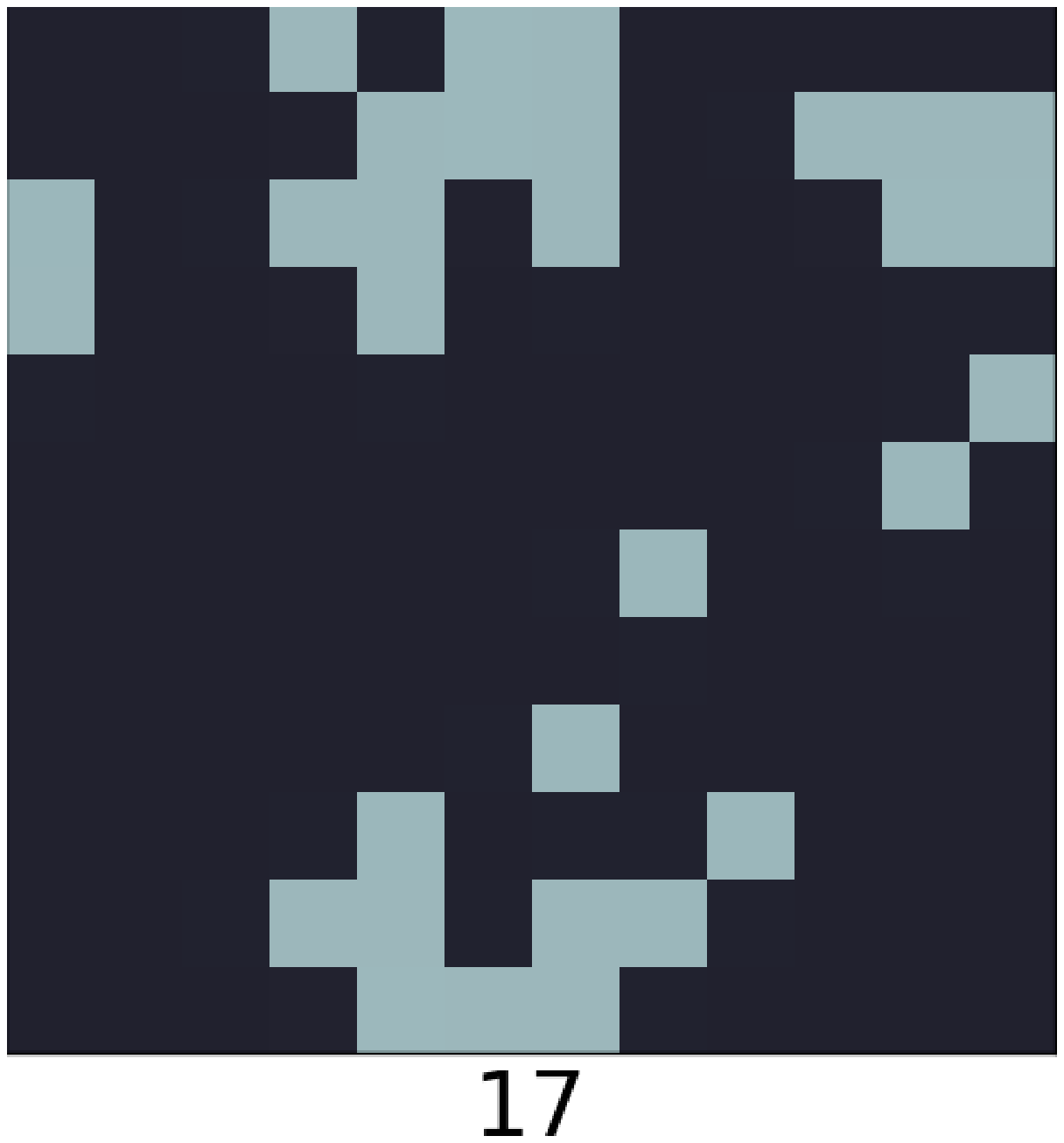}}\hspace{0.4em}
  \subfloat{\label{fig:snap14}\includegraphics[width=0.11\textwidth,height=0.8in]{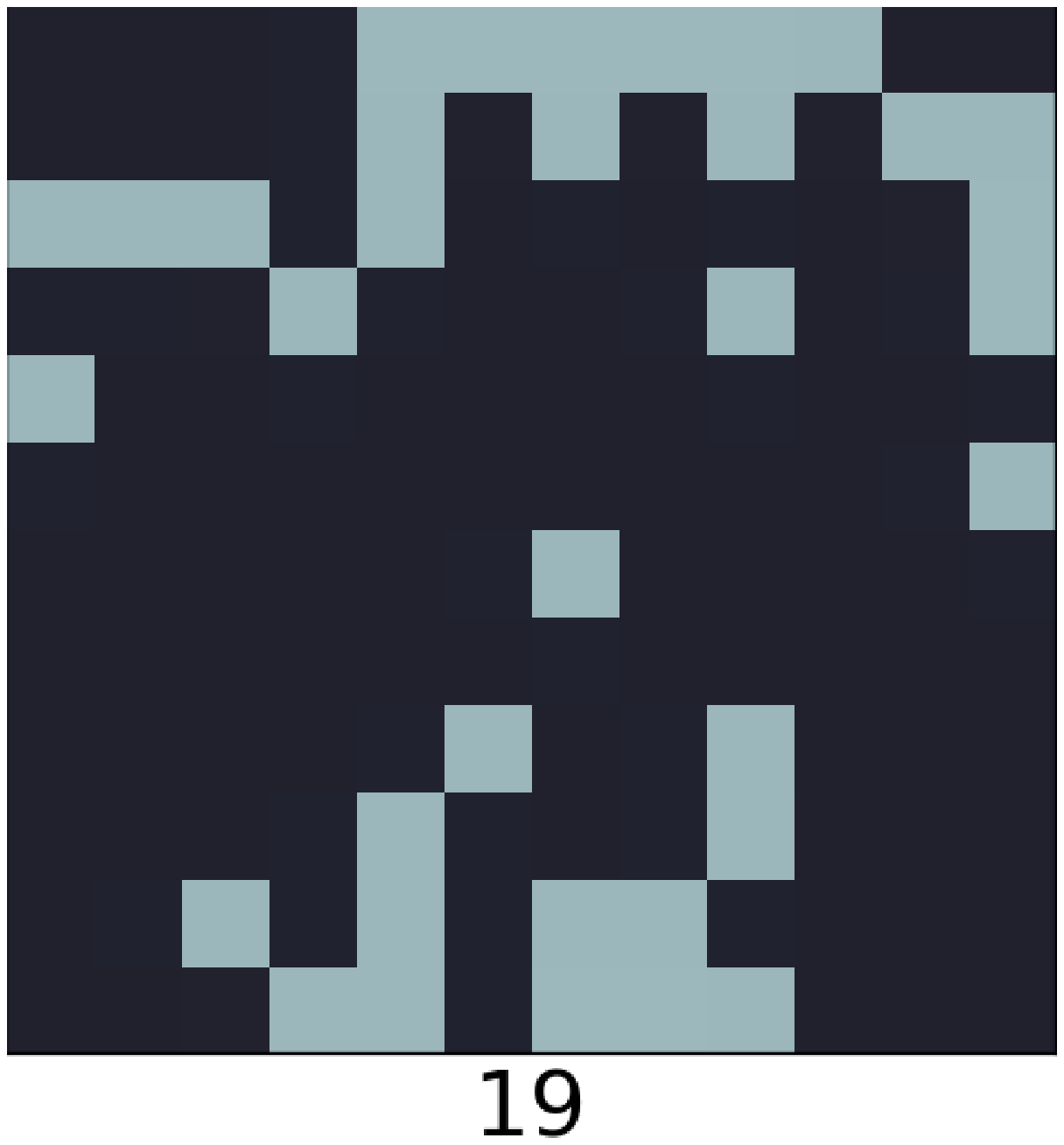}}\hspace{0.4em}
  \subfloat{\label{fig:snap15}\includegraphics[width=0.11\textwidth,height=0.8in]{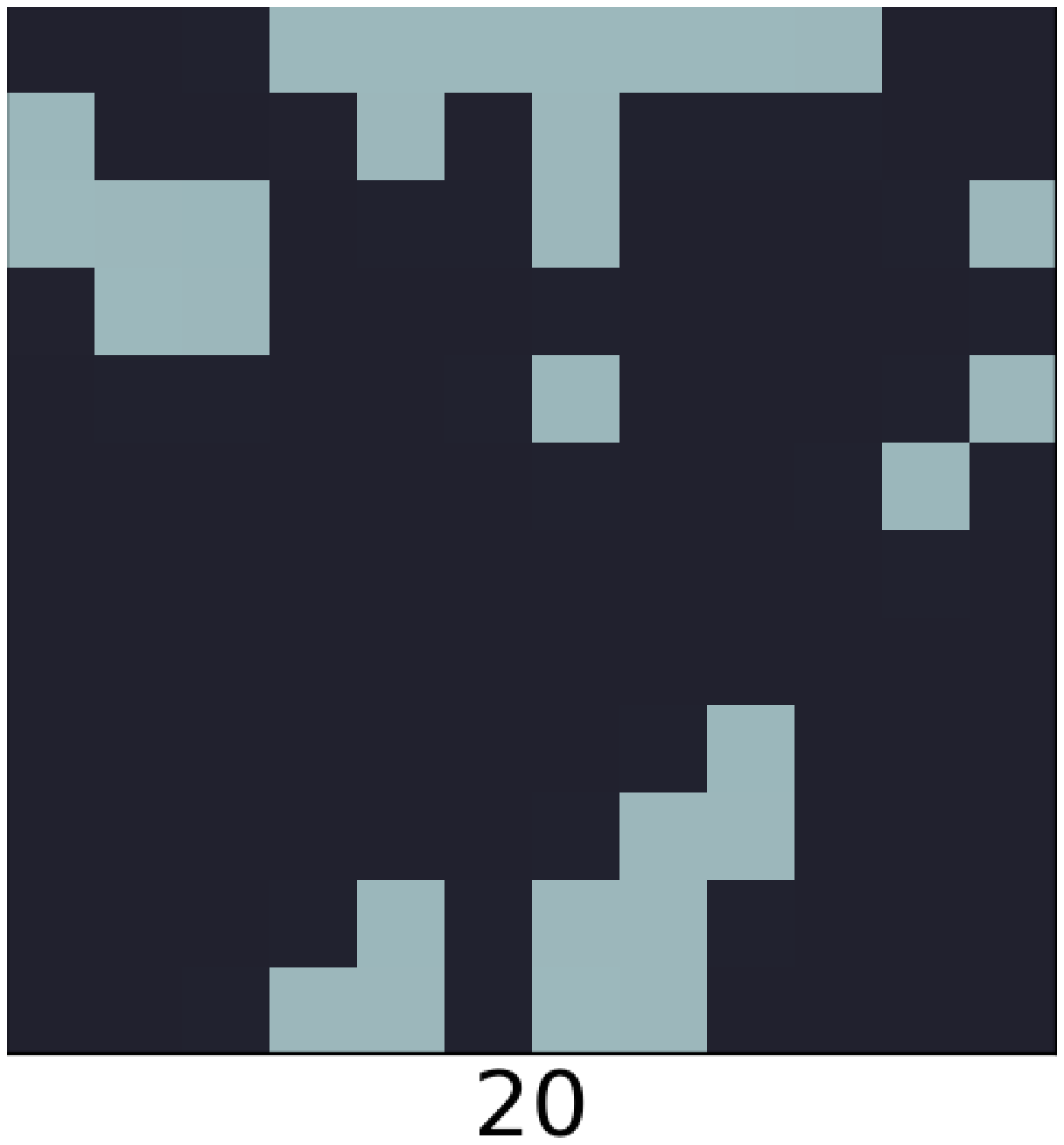}}\hspace{0.4em}\\

  \subfloat{\label{fig:snap16}\includegraphics[width=0.11\textwidth,height=0.8in]{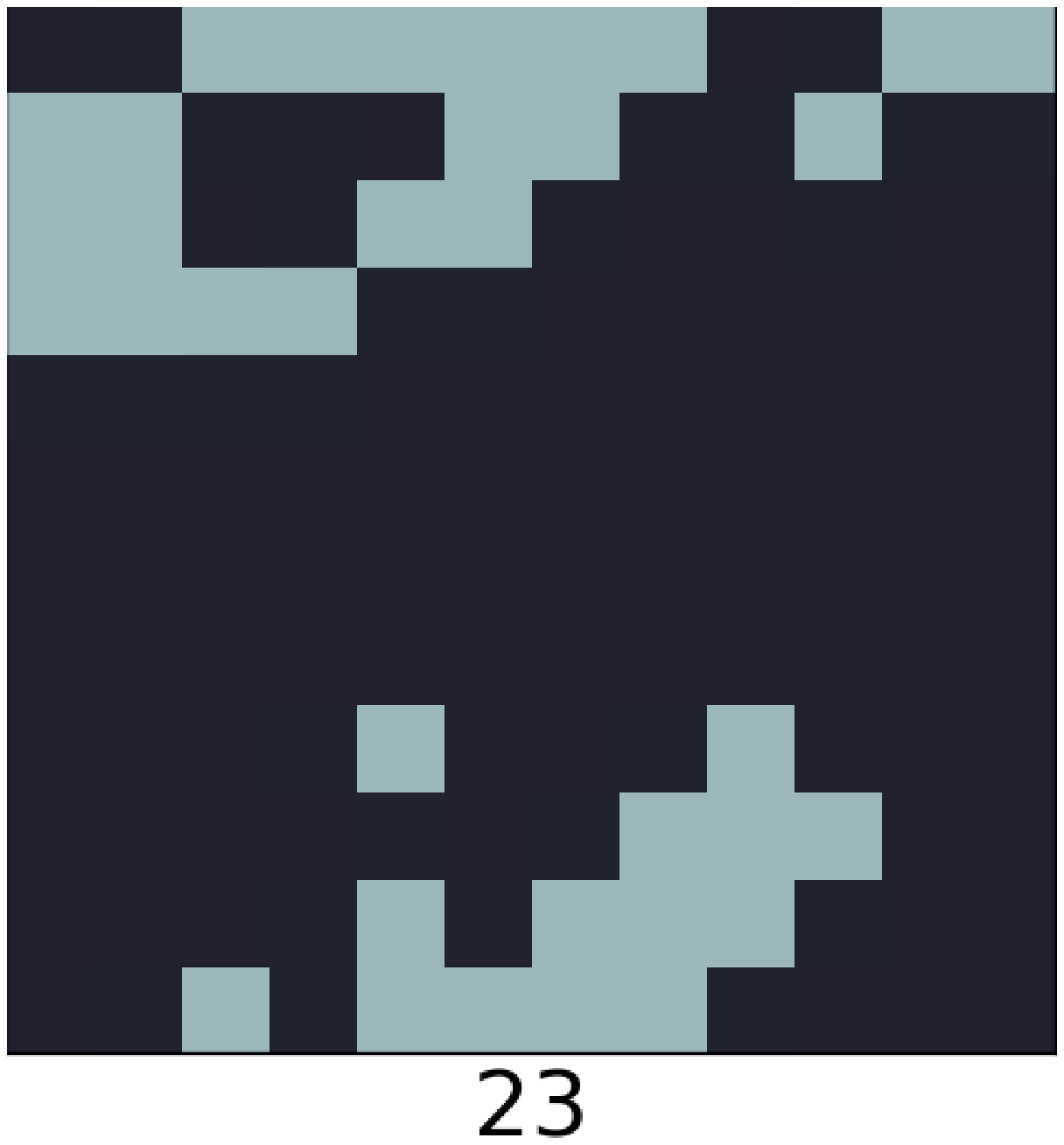}}\hspace{0.4em}
  \subfloat{\label{fig:snap17}\includegraphics[width=0.11\textwidth,height=0.8in]{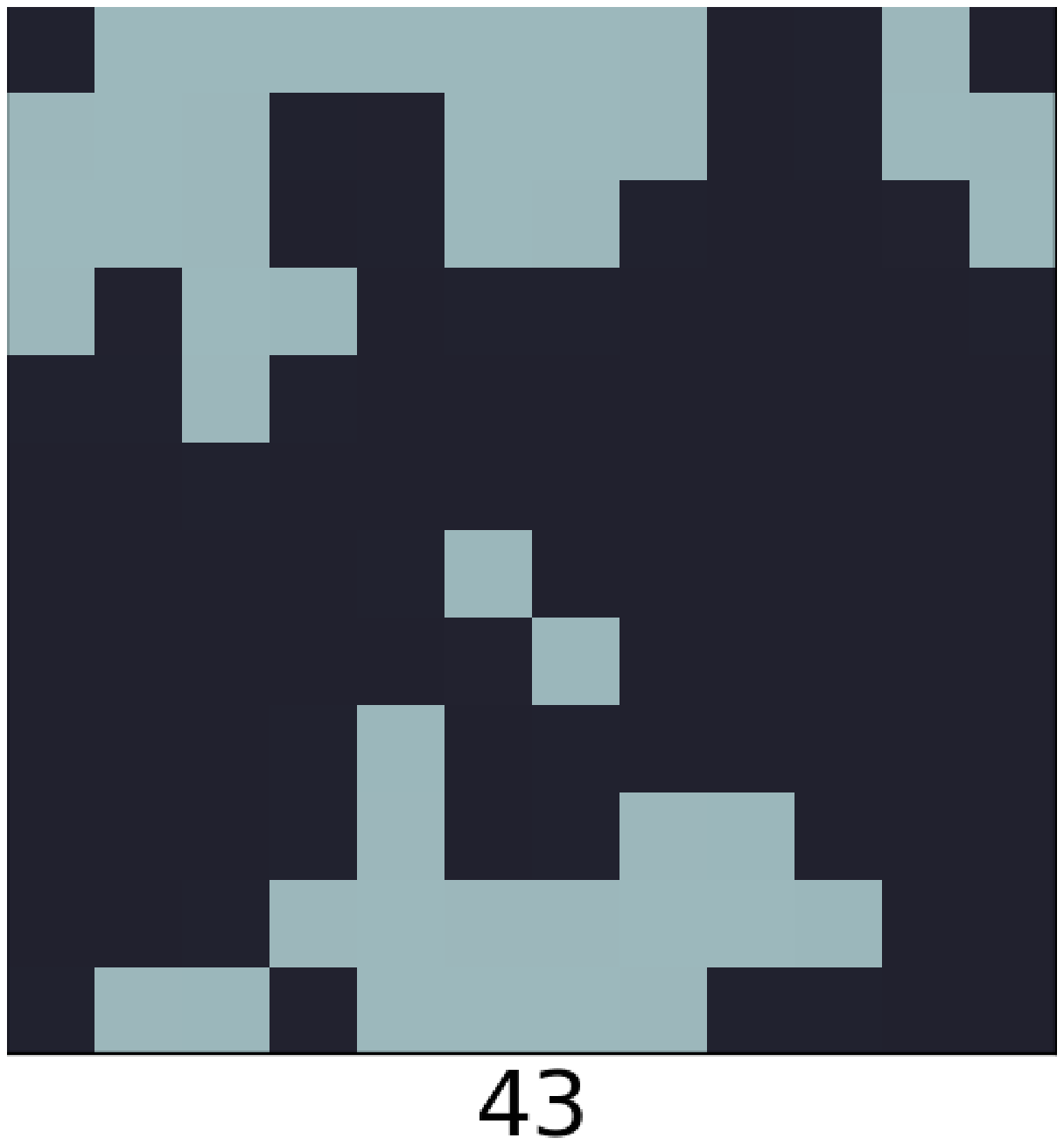}}\hspace{0.4em}
  \subfloat{\label{fig:snap18}\includegraphics[width=0.11\textwidth,height=0.8in]{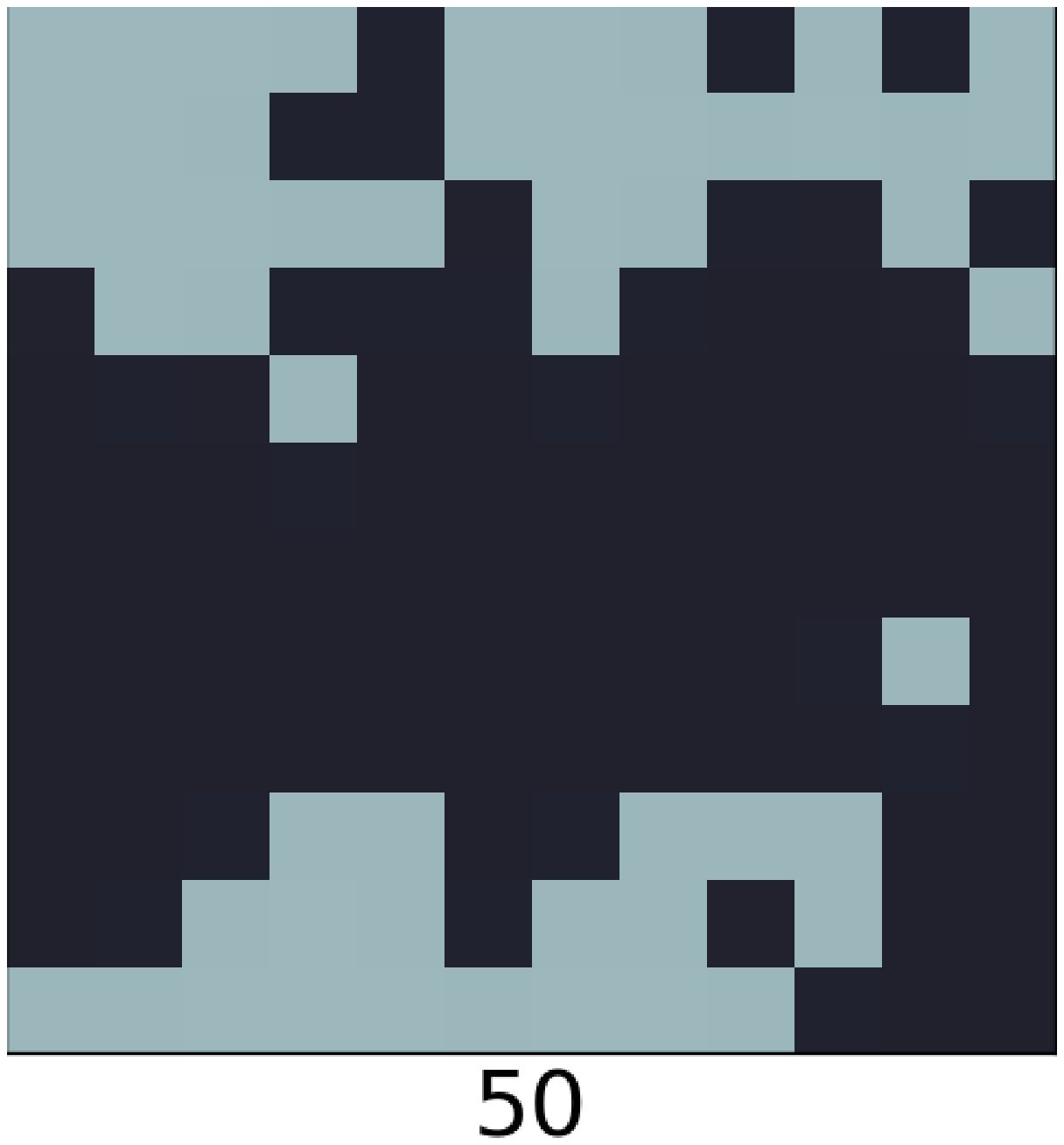}}\hspace{0.4em}
  \subfloat{\label{fig:snap19}\includegraphics[width=0.11\textwidth,height=0.8in]{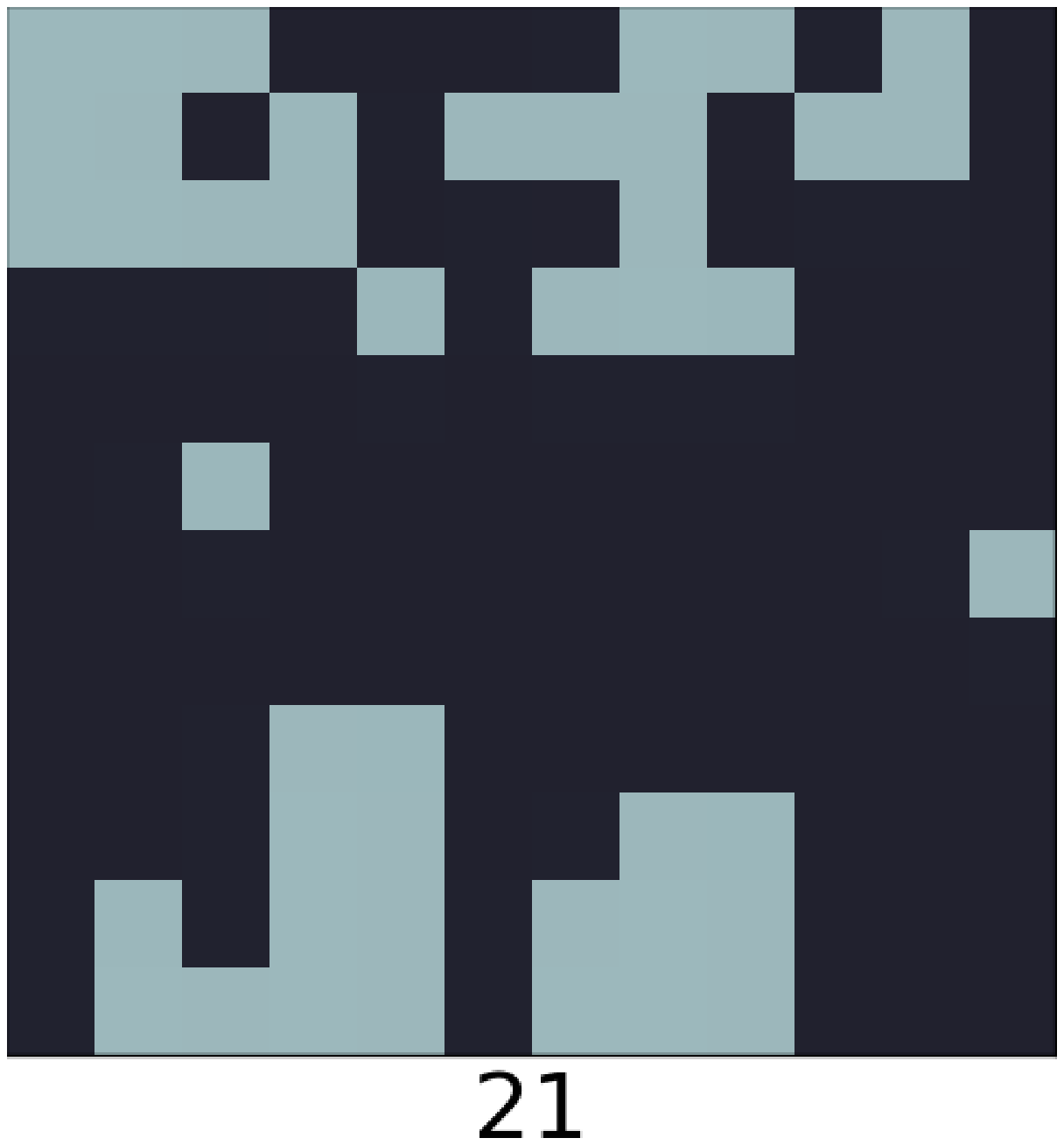}}\hspace{0.4em}
  \subfloat{\label{fig:snap20}\includegraphics[width=0.11\textwidth,height=0.8in]{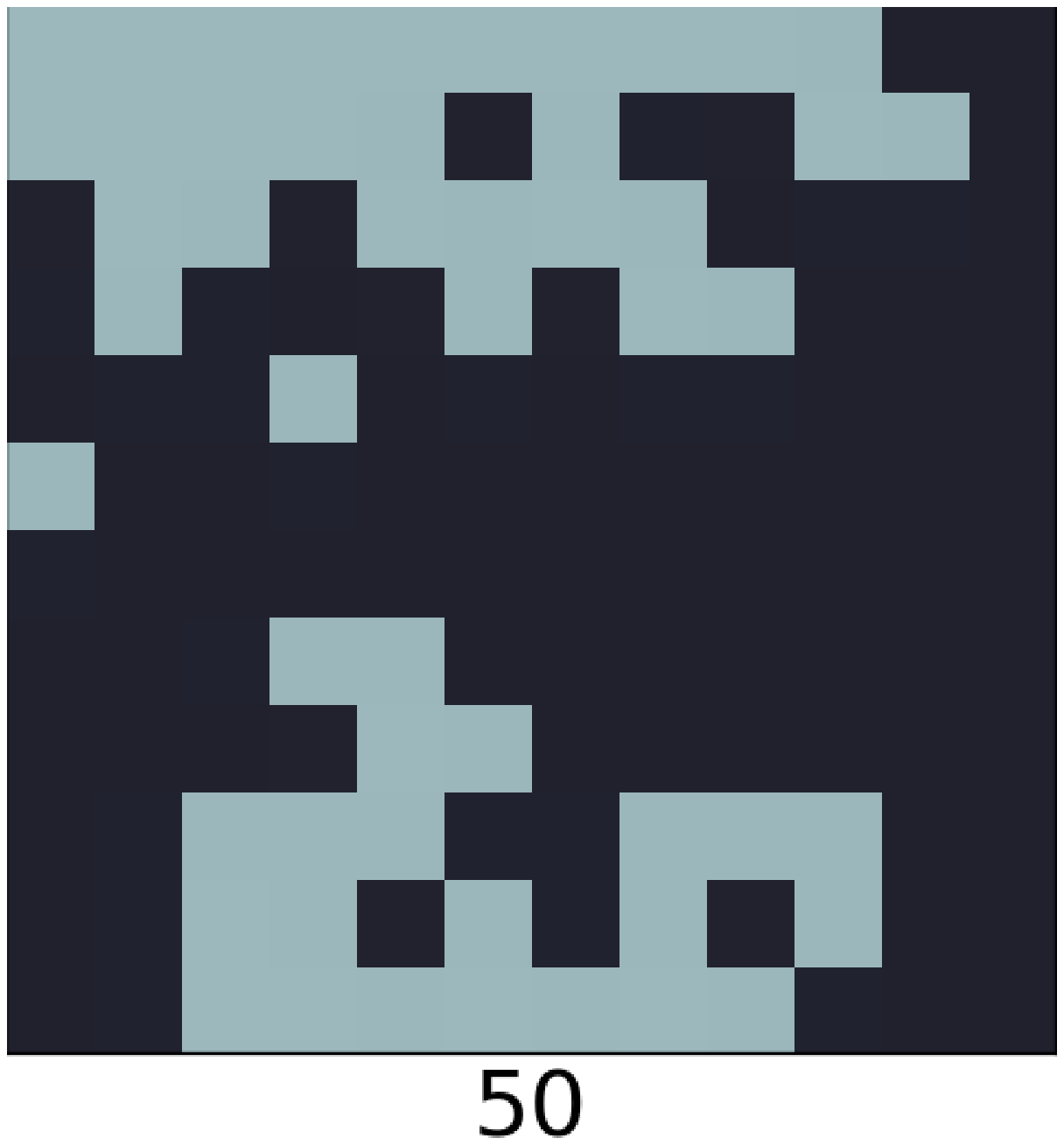}}\hspace{0.4em}
  \subfloat{\label{fig:snap21}\includegraphics[width=0.11\textwidth,height=0.8in]{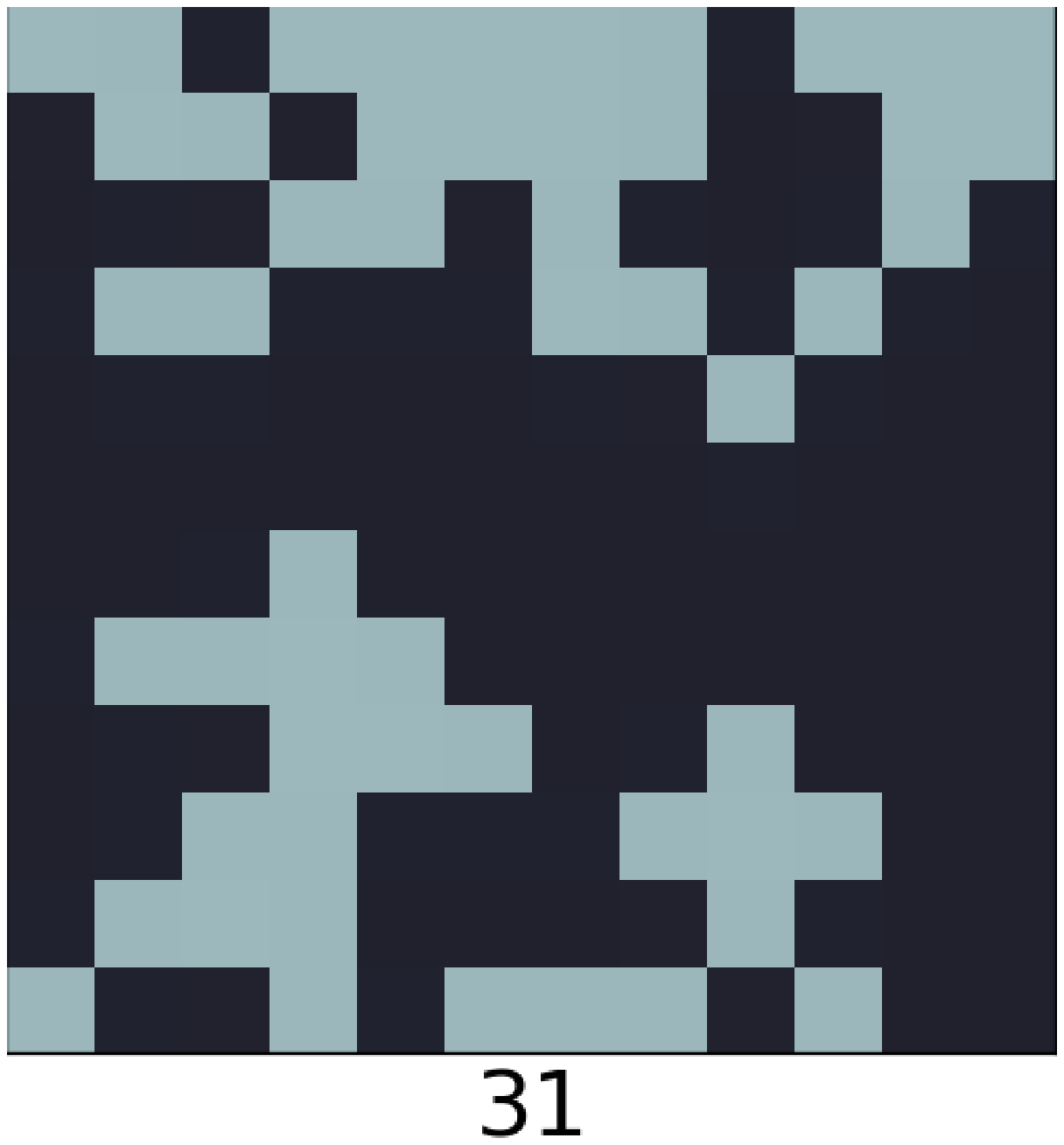}}\hspace{0.4em}
  \subfloat{\label{fig:snap22}\includegraphics[width=0.11\textwidth,height=0.8in]{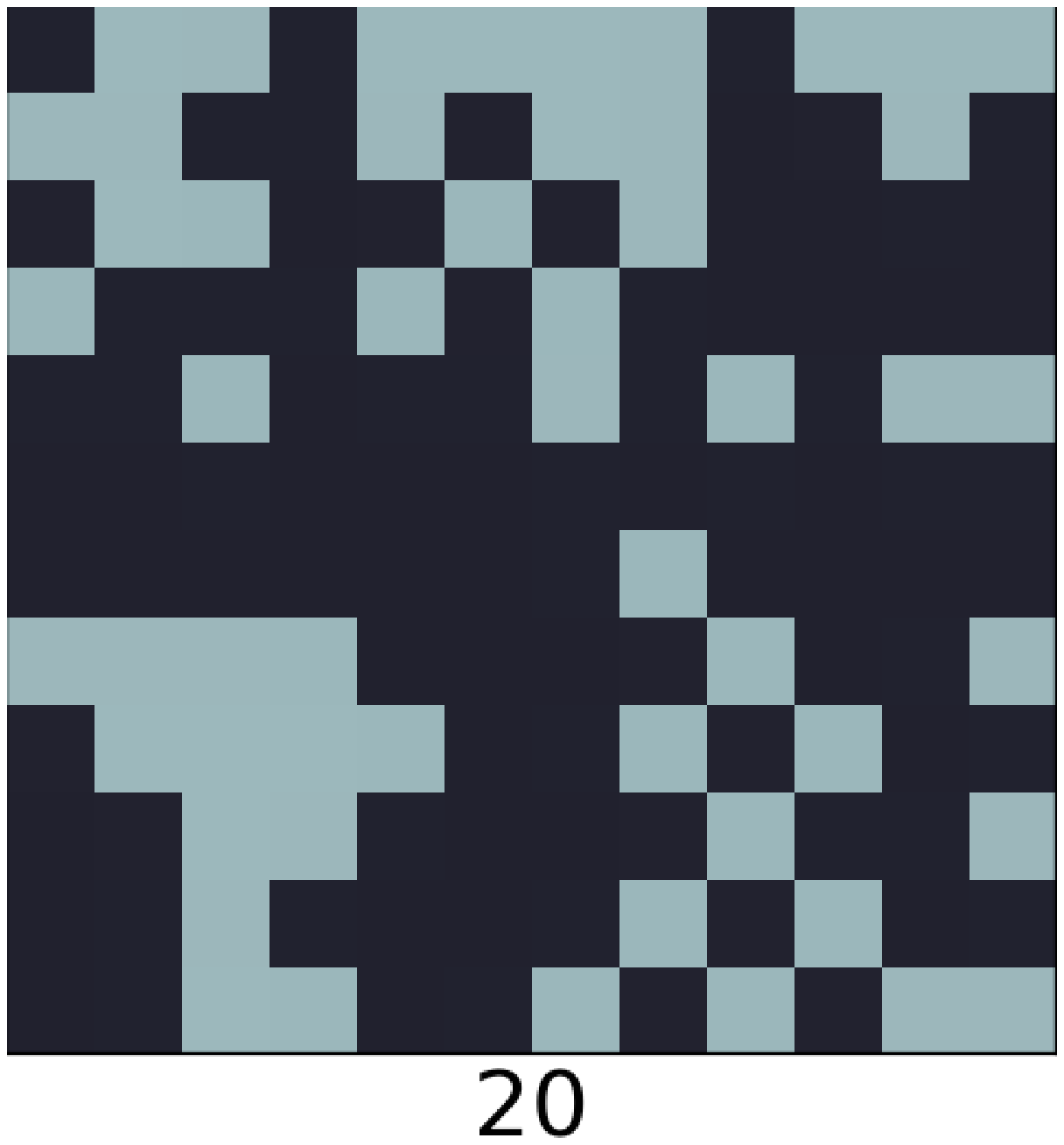}}\hspace{0.4em}
  \subfloat{\label{fig:snap23}\includegraphics[width=0.11\textwidth,height=0.8in]{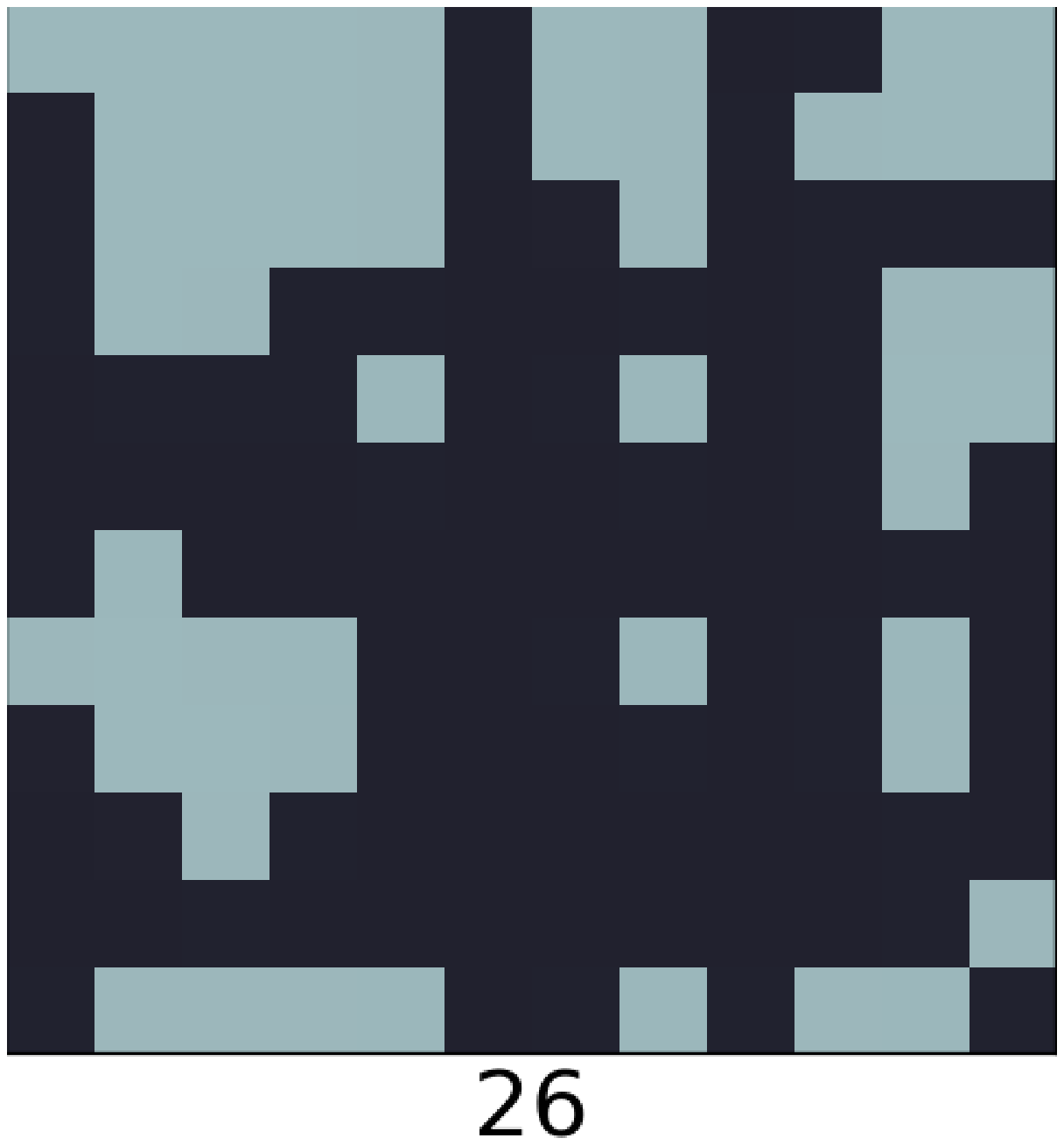}}\hspace{0.4em}\\

  \subfloat{\label{fig:snap24}\includegraphics[width=0.11\textwidth,height=0.8in]{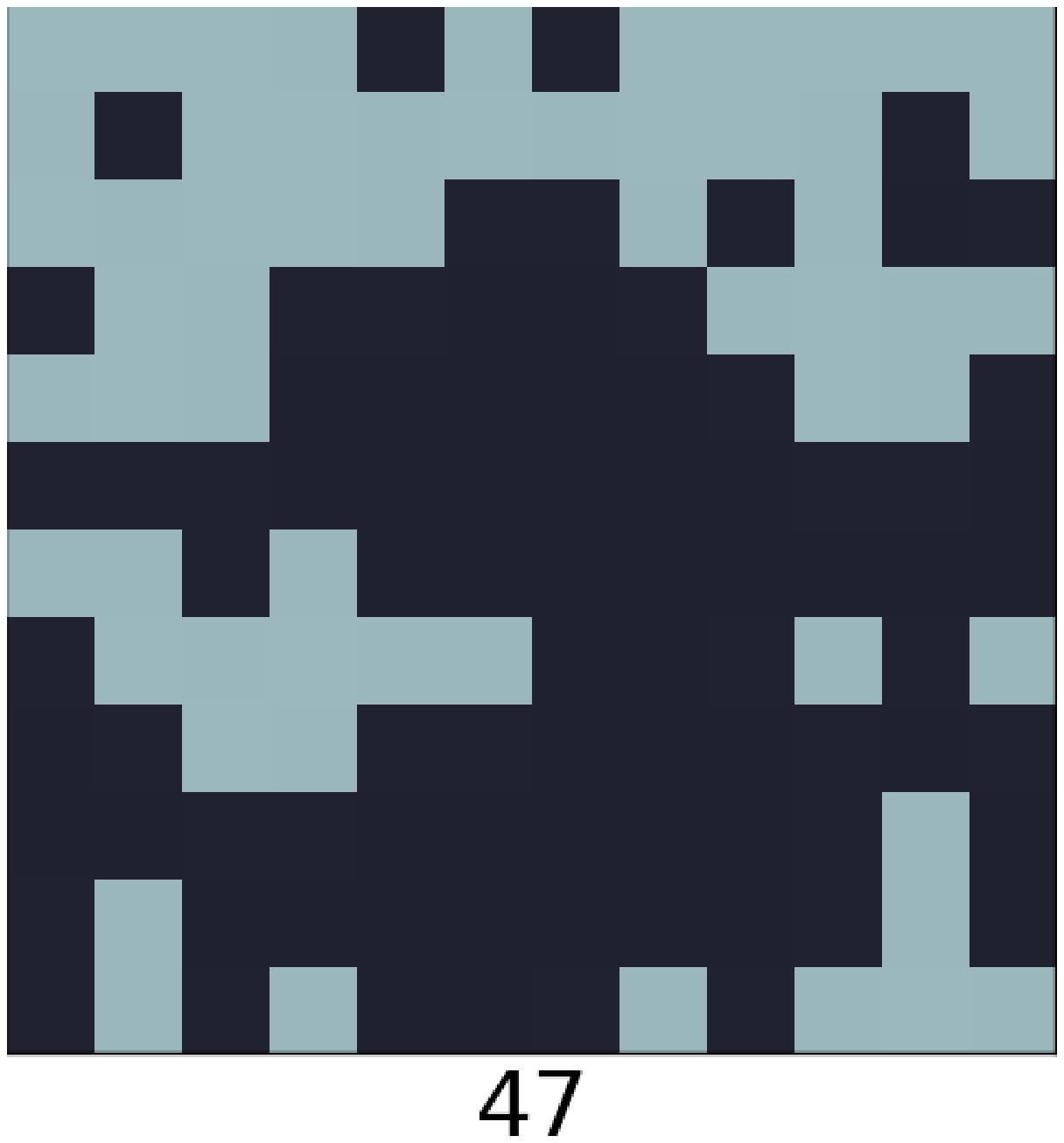}}\hspace{0.4em}
  \subfloat{\label{fig:snap25}\includegraphics[width=0.11\textwidth,height=0.8in]{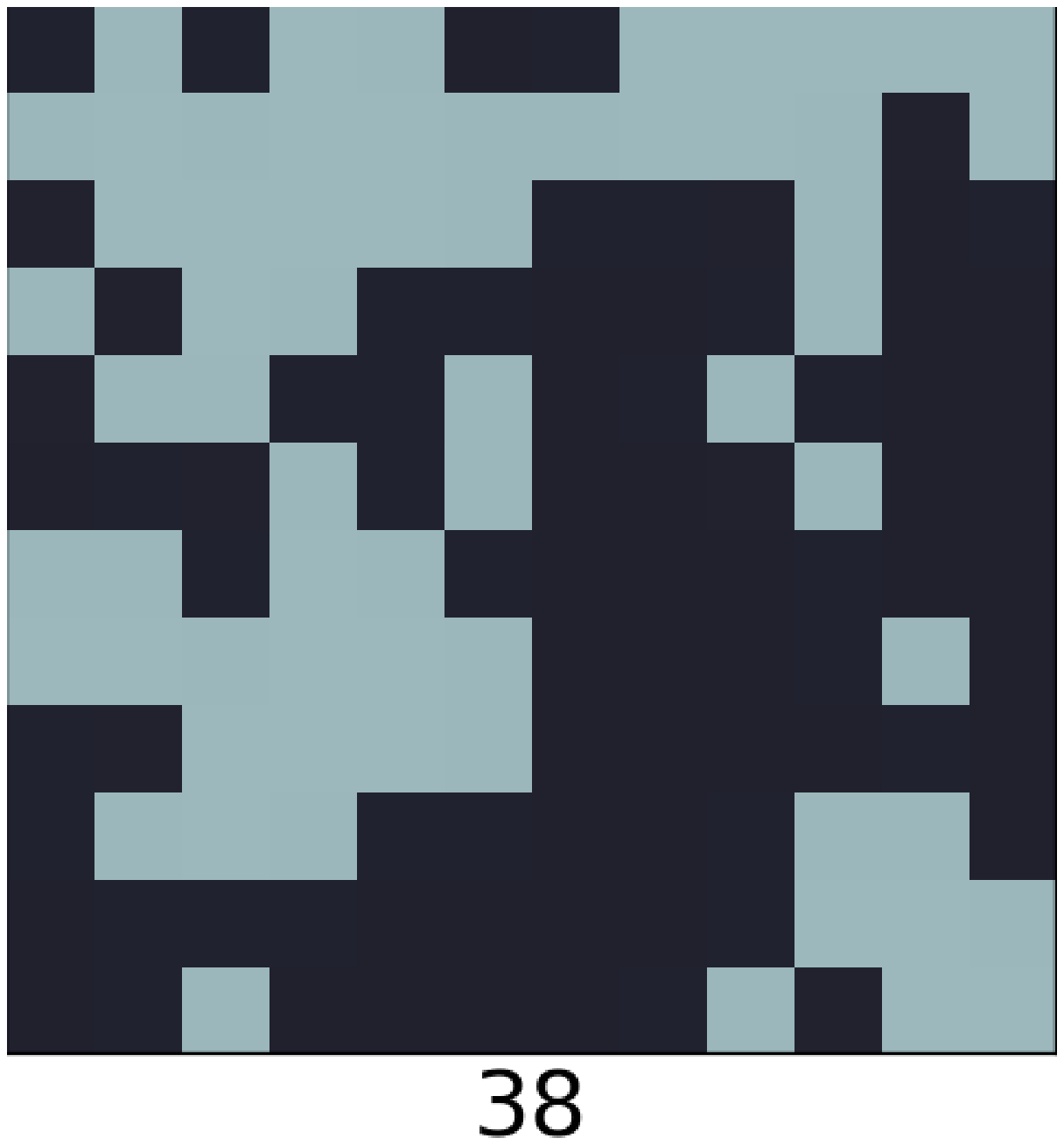}}\hspace{0.4em}
  \subfloat{\label{fig:snap26}\includegraphics[width=0.11\textwidth,height=0.8in]{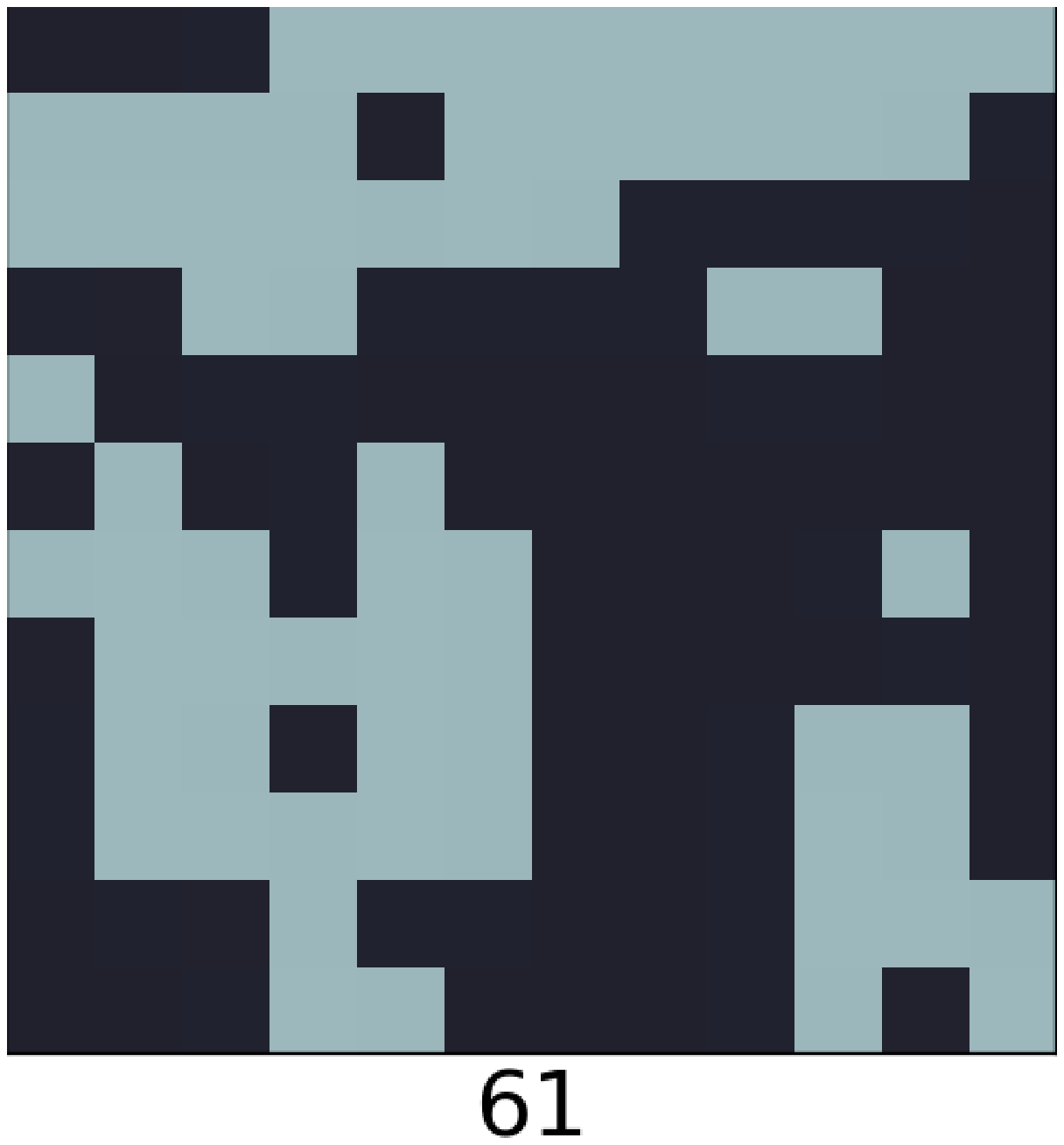}}\hspace{0.4em}
  \subfloat{\label{fig:snap27}\includegraphics[width=0.11\textwidth,height=0.8in]{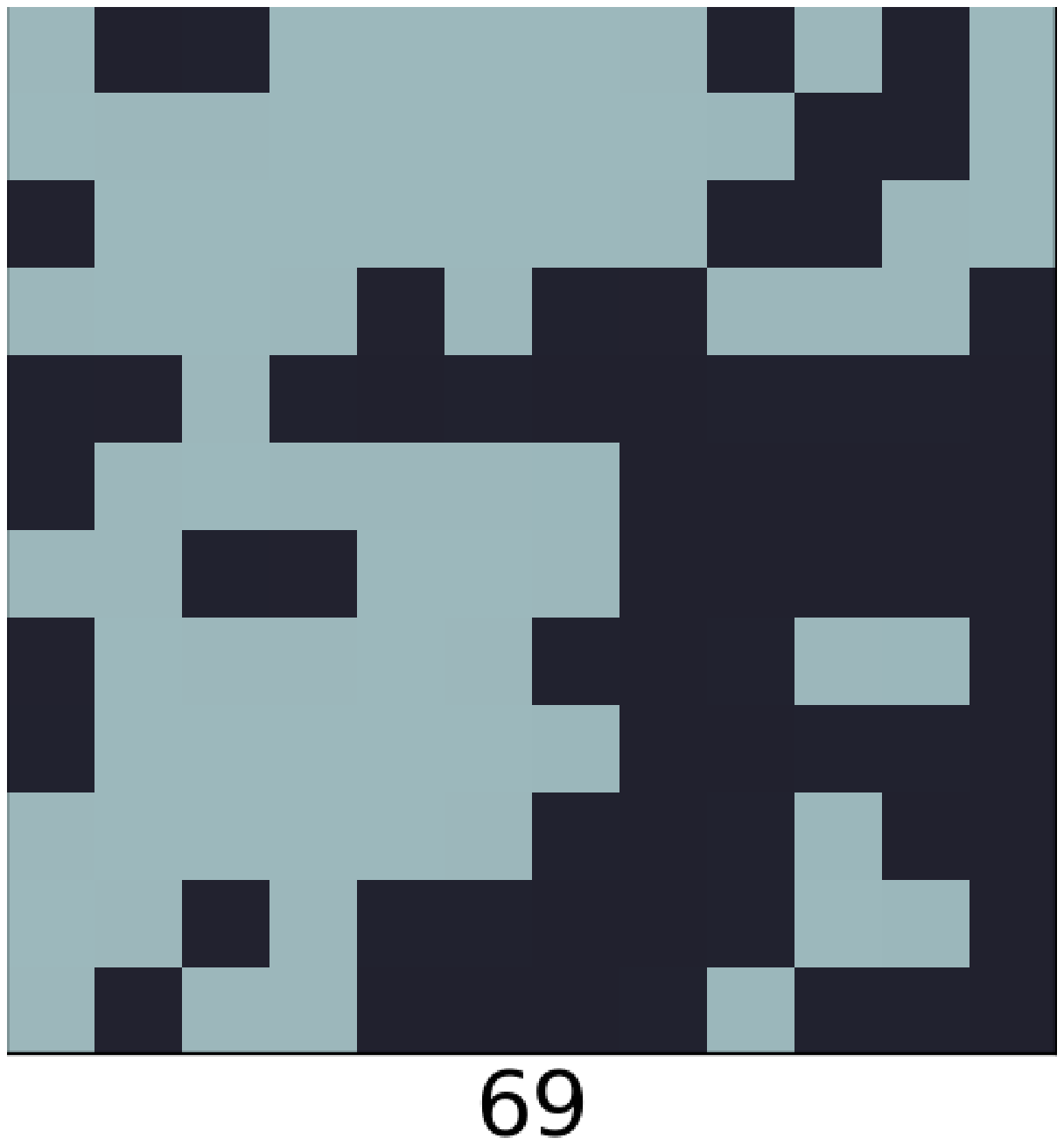}}\hspace{0.4em}
  \subfloat{\label{fig:snap28}\includegraphics[width=0.11\textwidth,height=0.8in]{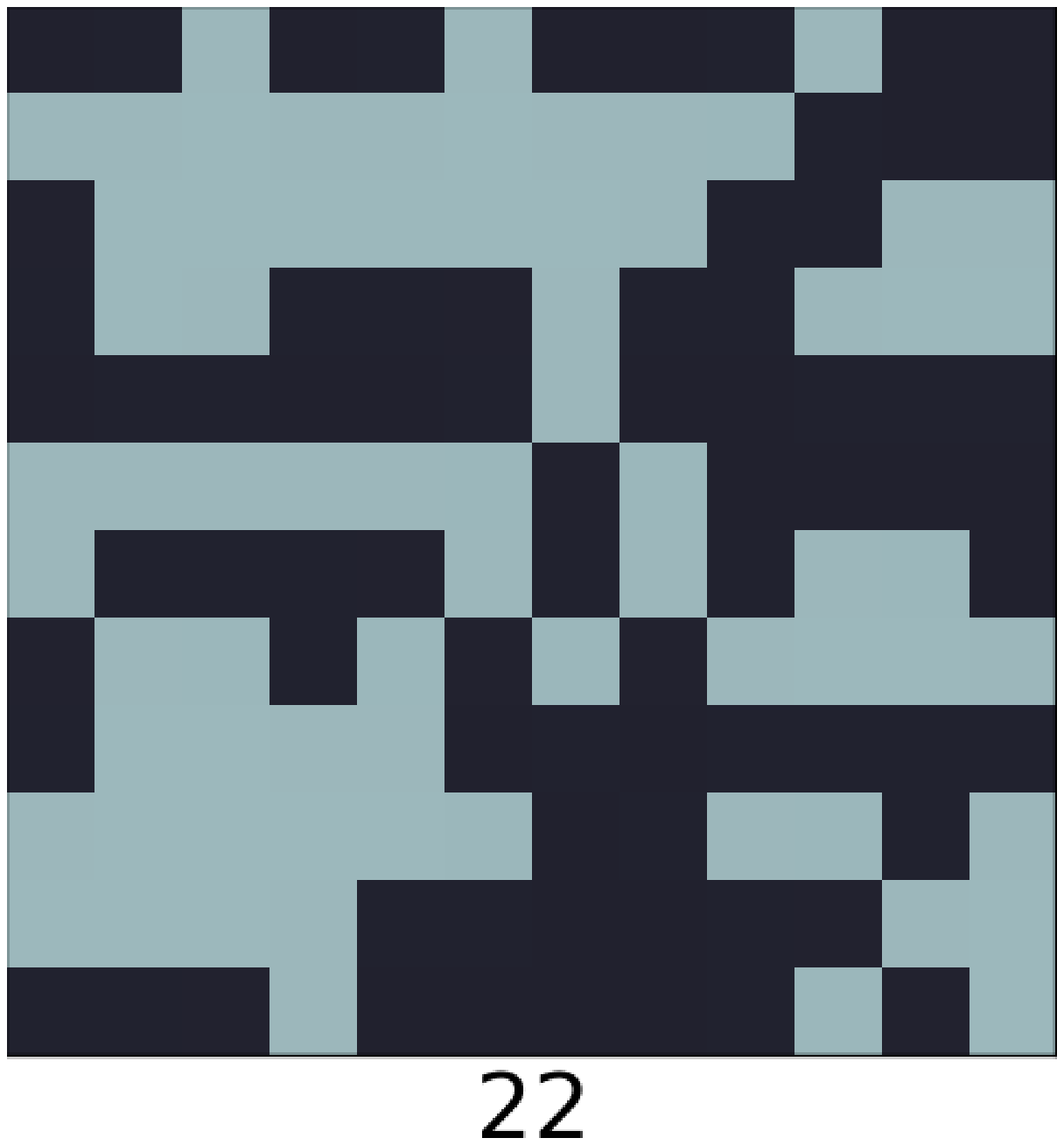}}\hspace{0.4em}
  \subfloat{\label{fig:snap29}\includegraphics[width=0.11\textwidth,height=0.8in]{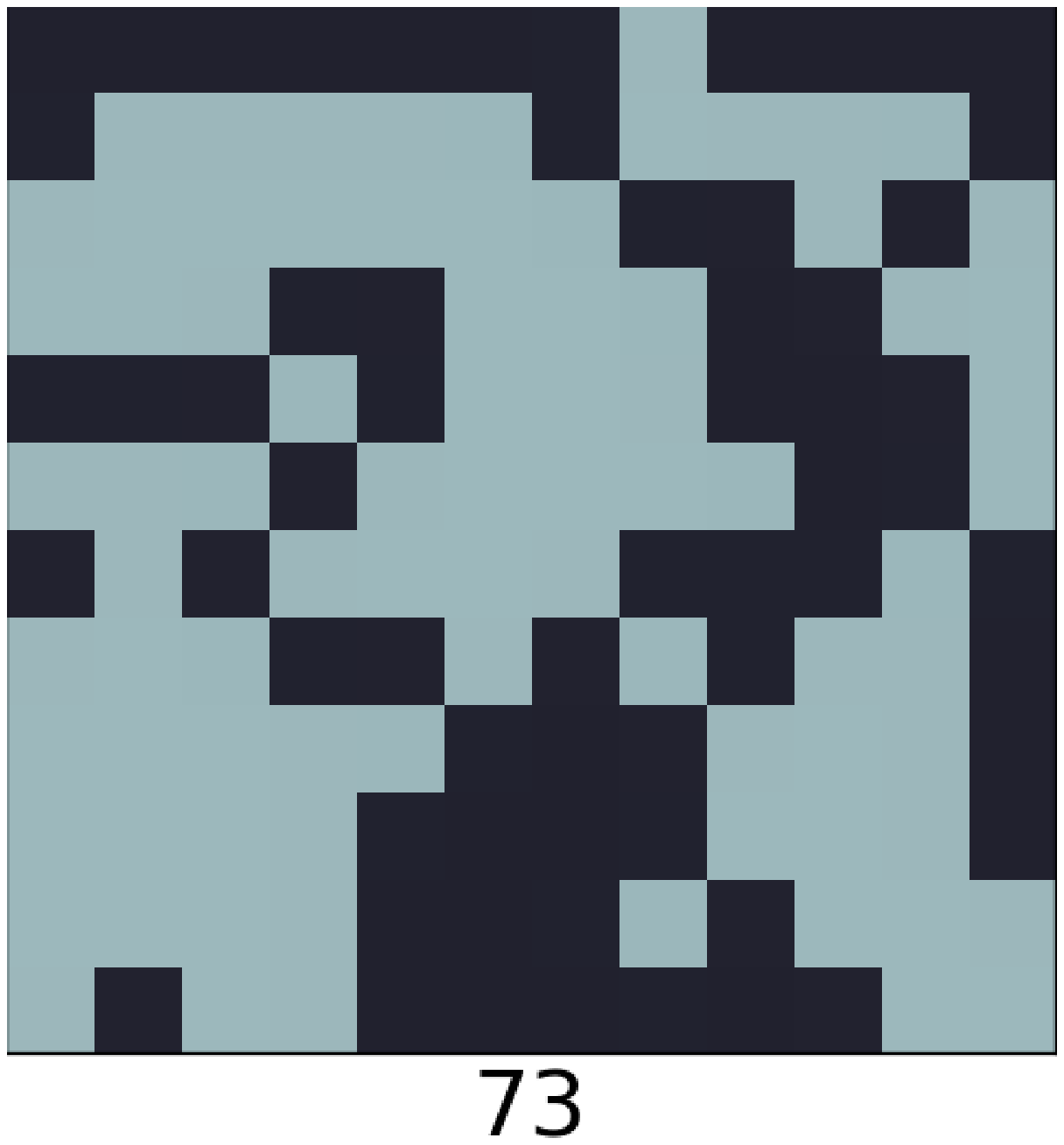}}\hspace{0.4em}
  \subfloat{\label{fig:snap30}\includegraphics[width=0.11\textwidth,height=0.8in]{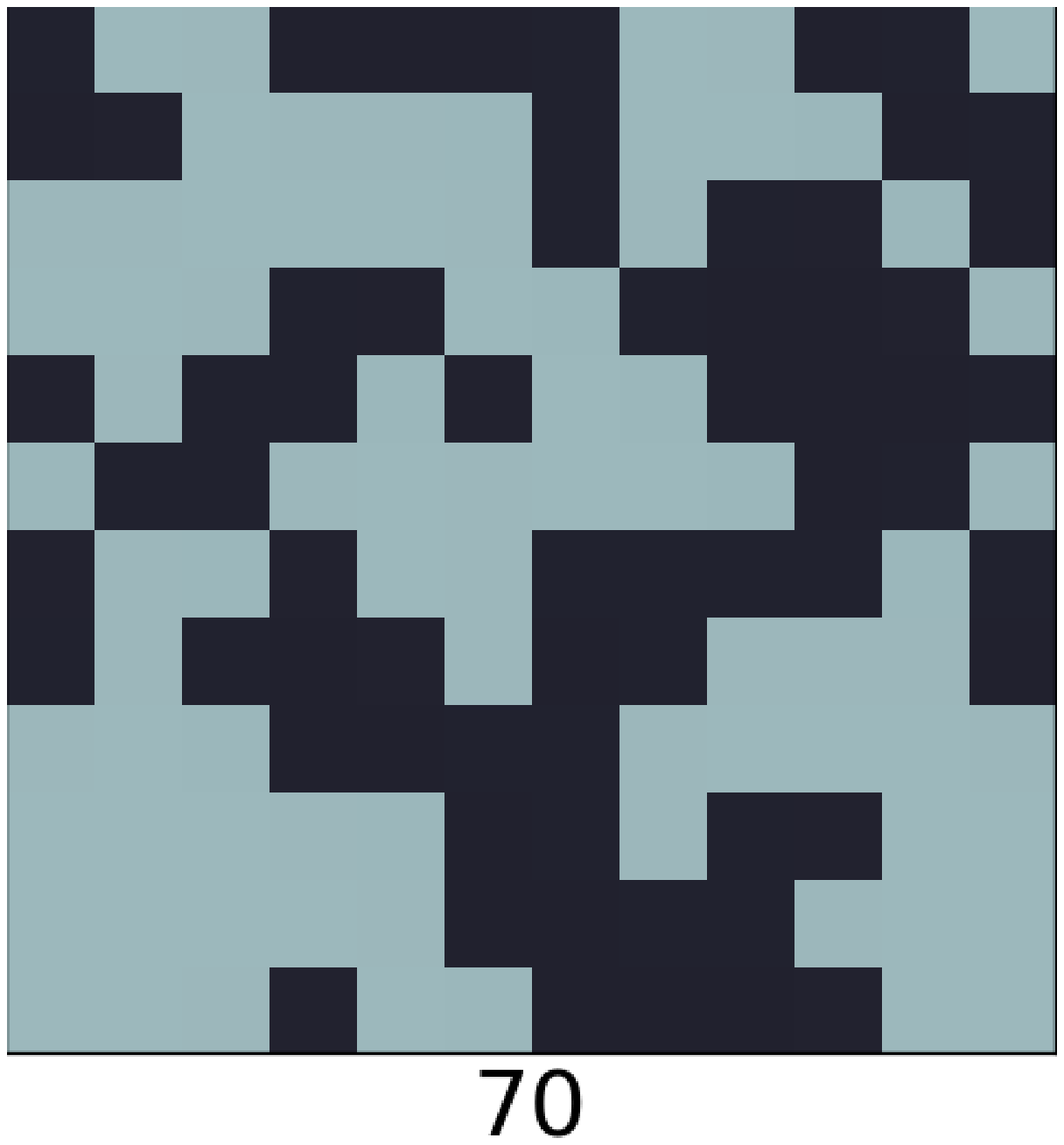}}\hspace{0.4em}
  \subfloat{\label{fig:snap31}\includegraphics[width=0.11\textwidth,height=0.8in]{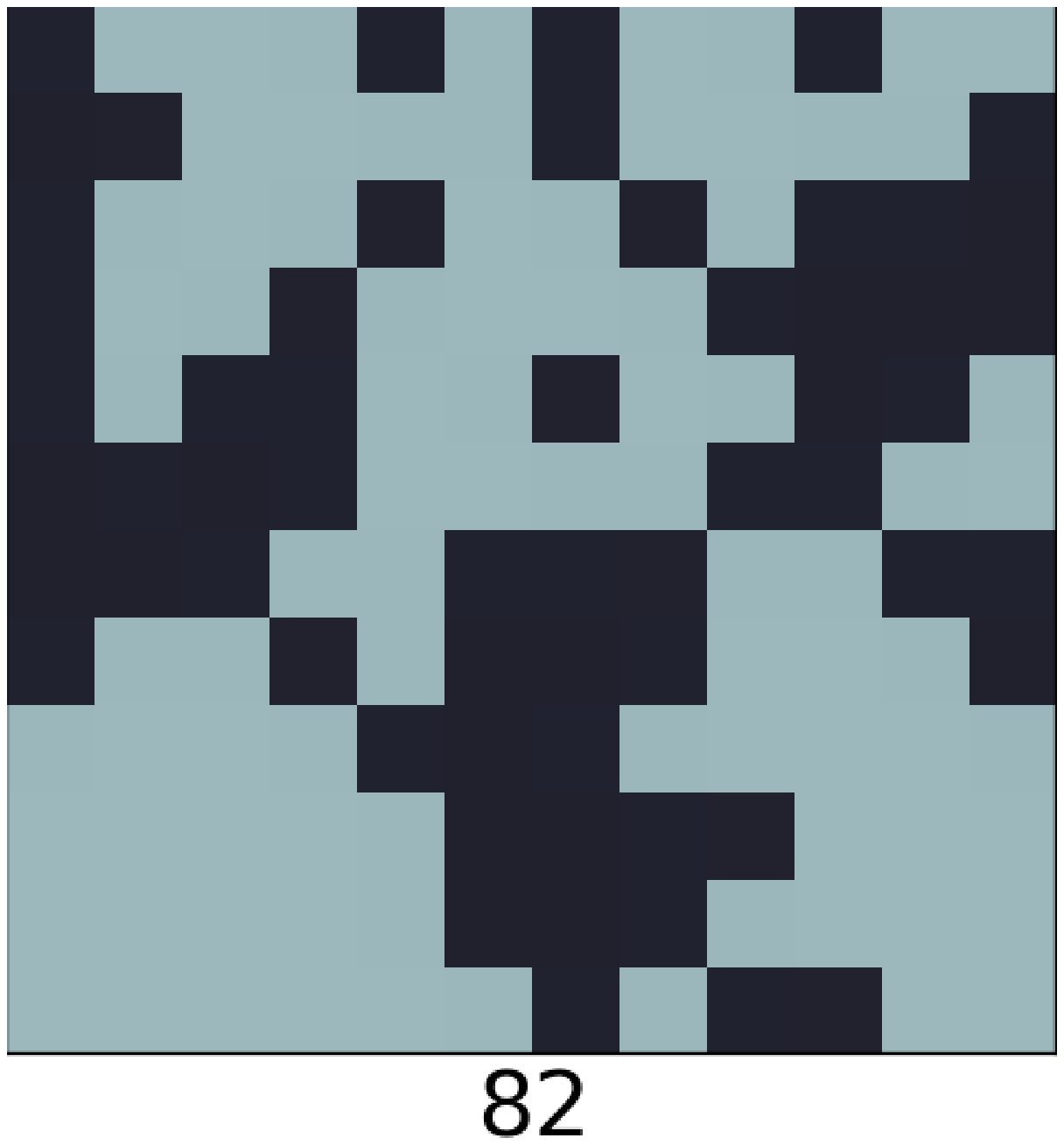}}\hspace{0.4em}\\

  \subfloat{\label{fig:snap32}\includegraphics[width=0.11\textwidth,height=0.8in]{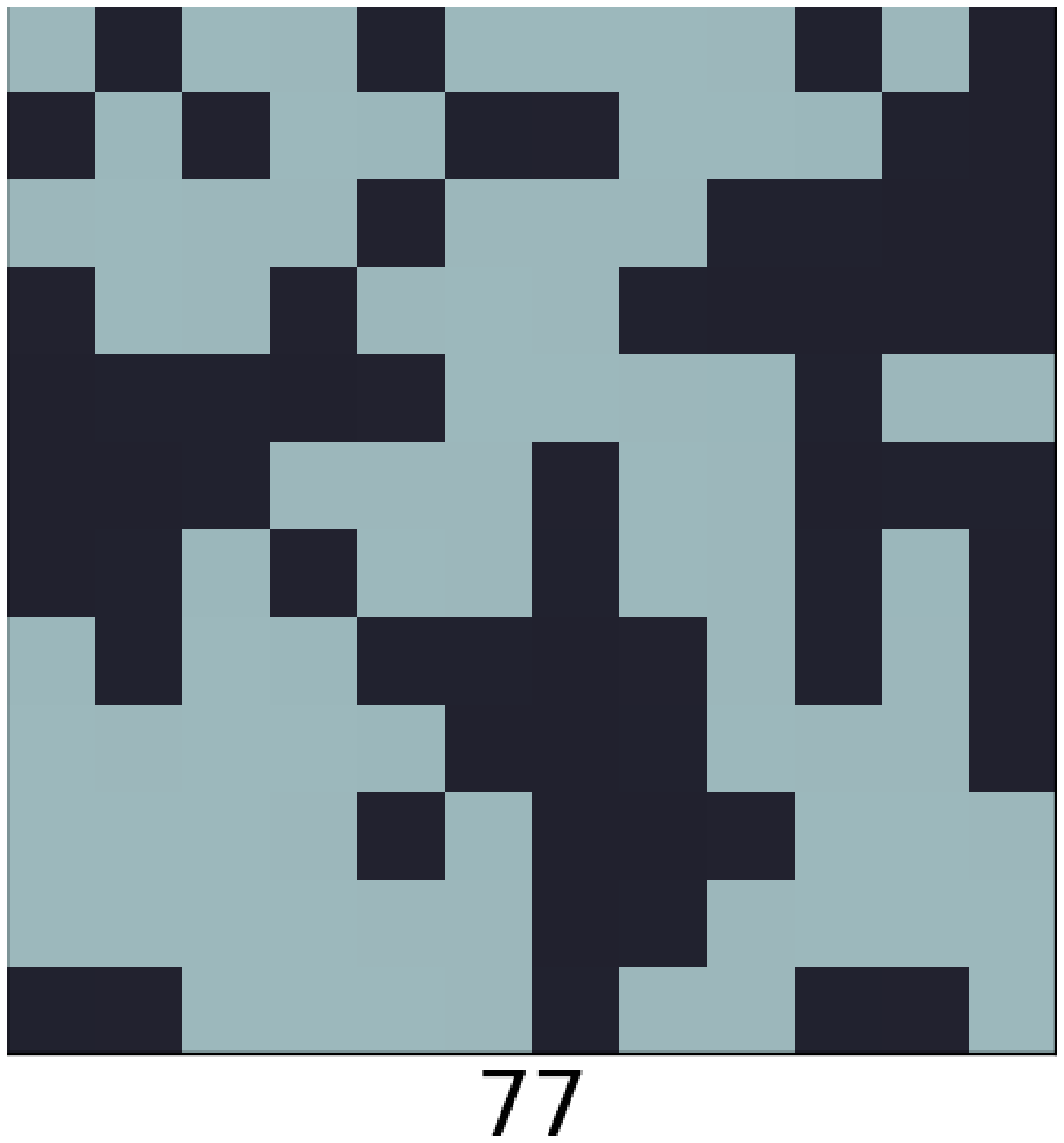}}\hspace{0.4em}
  \subfloat{\label{fig:snap33}\includegraphics[width=0.11\textwidth,height=0.8in]{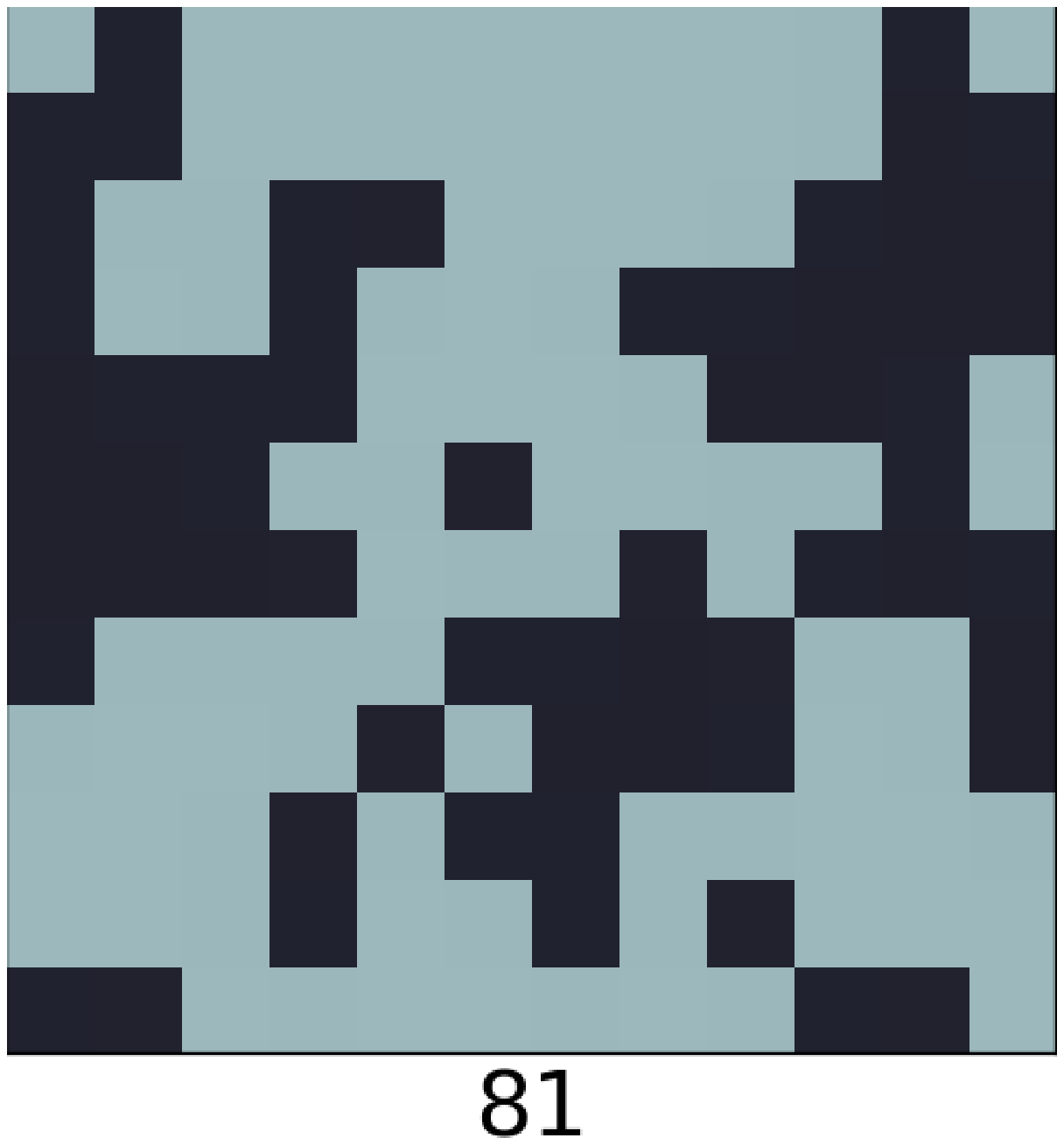}}\hspace{0.4em}
  \subfloat{\label{fig:snap34}\includegraphics[width=0.11\textwidth,height=0.8in]{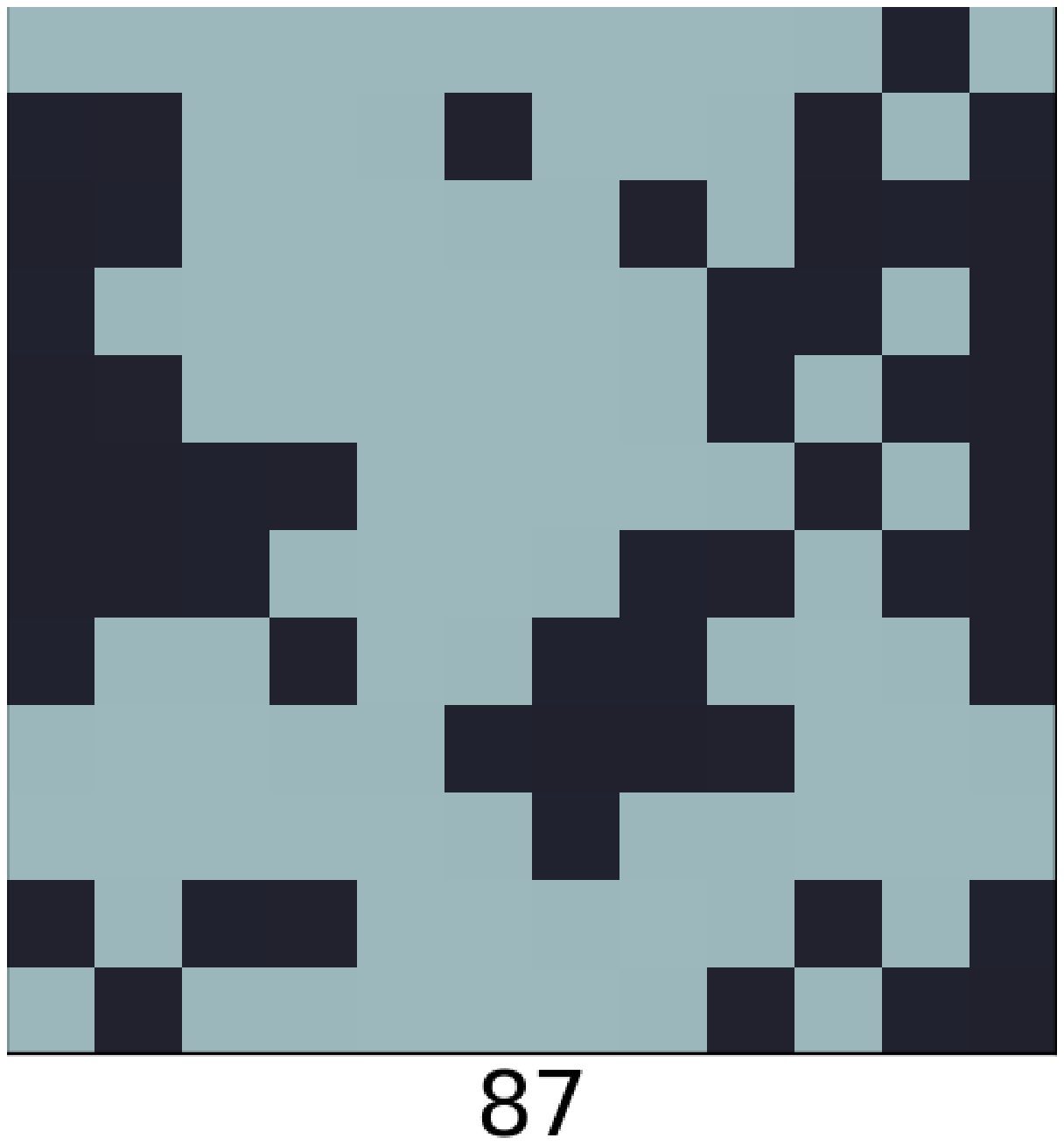}}\hspace{0.4em}
  \subfloat{\label{fig:snap35}\includegraphics[width=0.11\textwidth,height=0.8in]{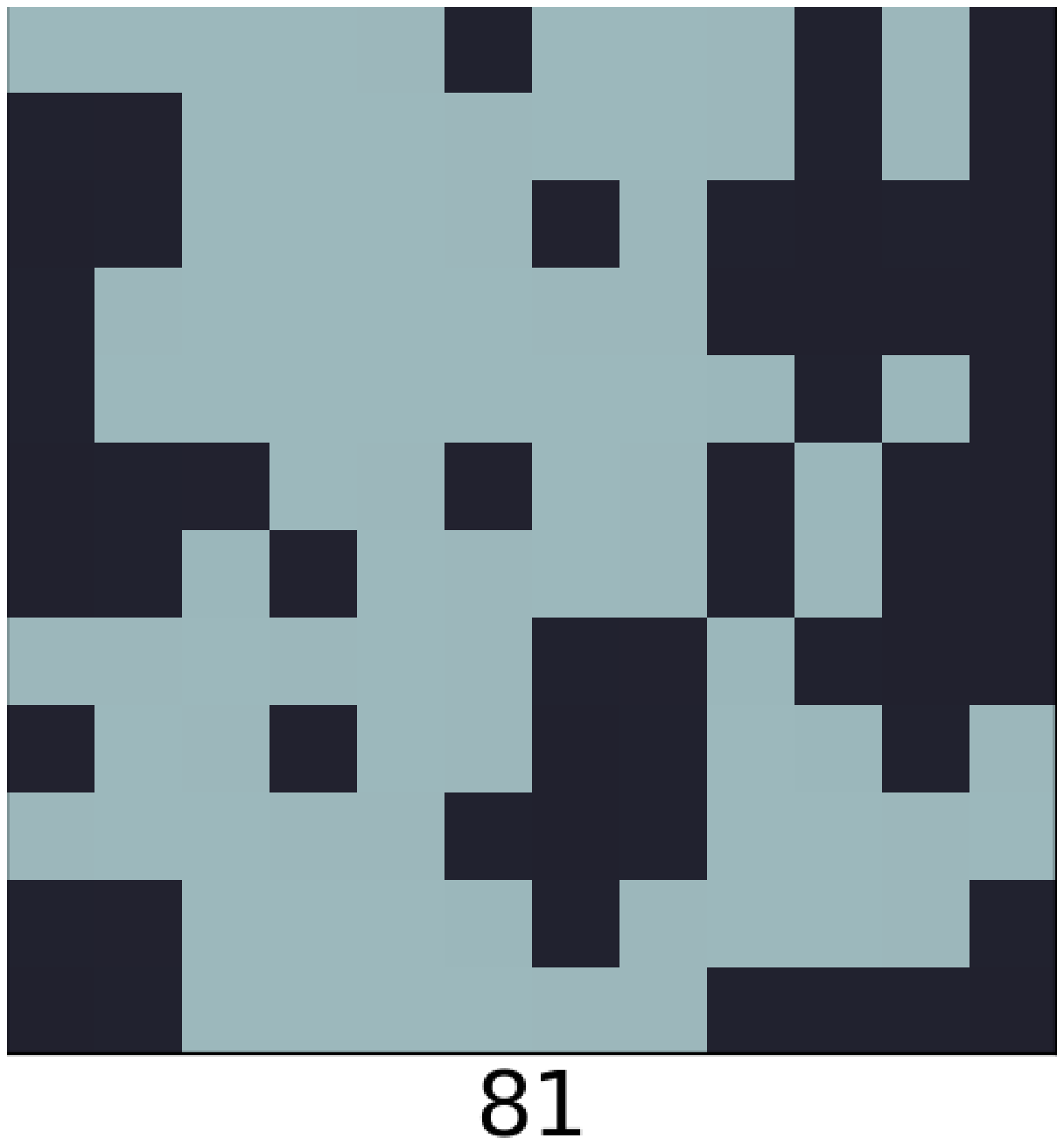}}\hspace{0.4em}
  \subfloat{\label{fig:snap36}\includegraphics[width=0.11\textwidth,height=0.8in]{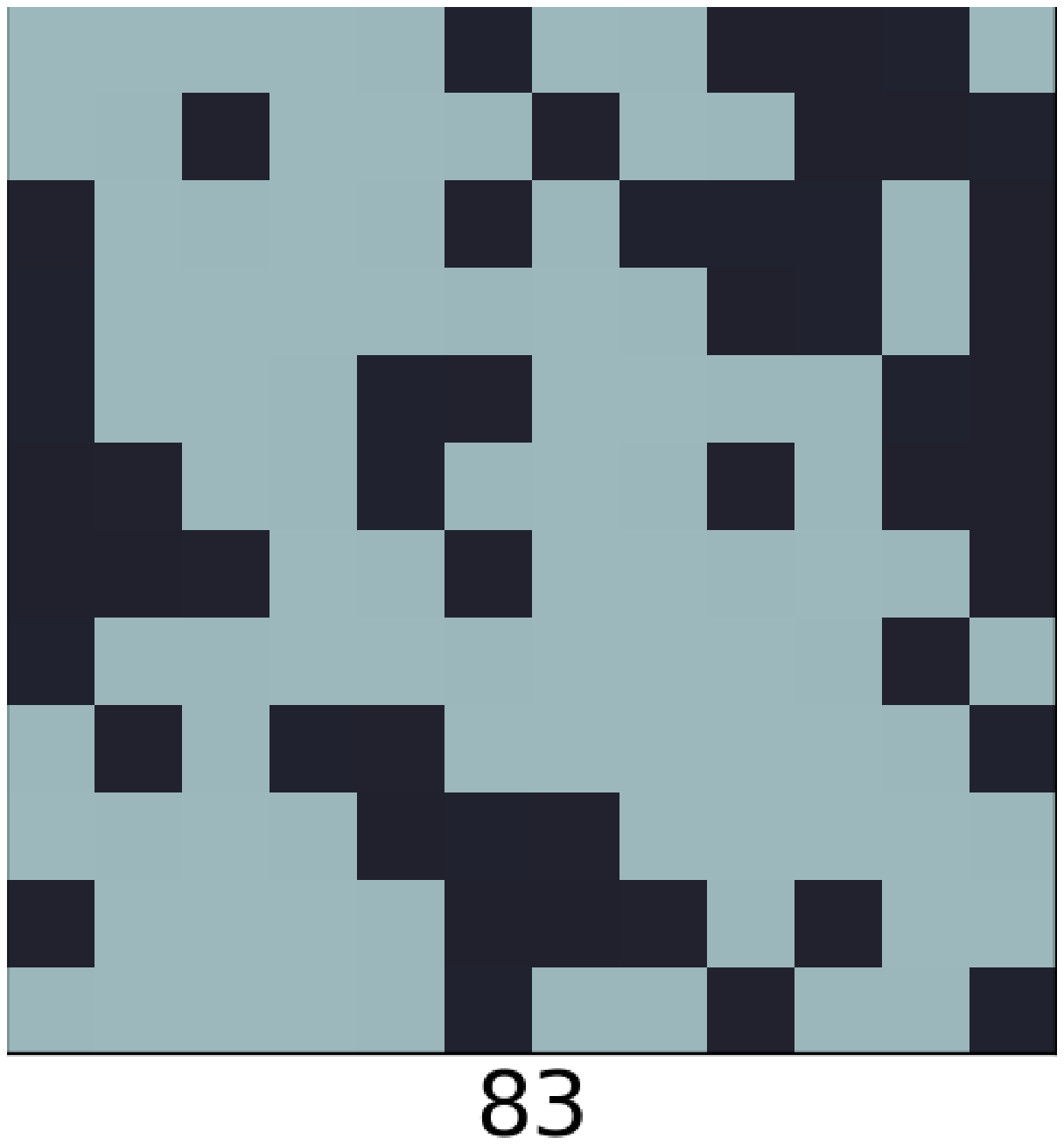}}\hspace{0.4em}
  \subfloat{\label{fig:snap37}\includegraphics[width=0.11\textwidth,height=0.8in]{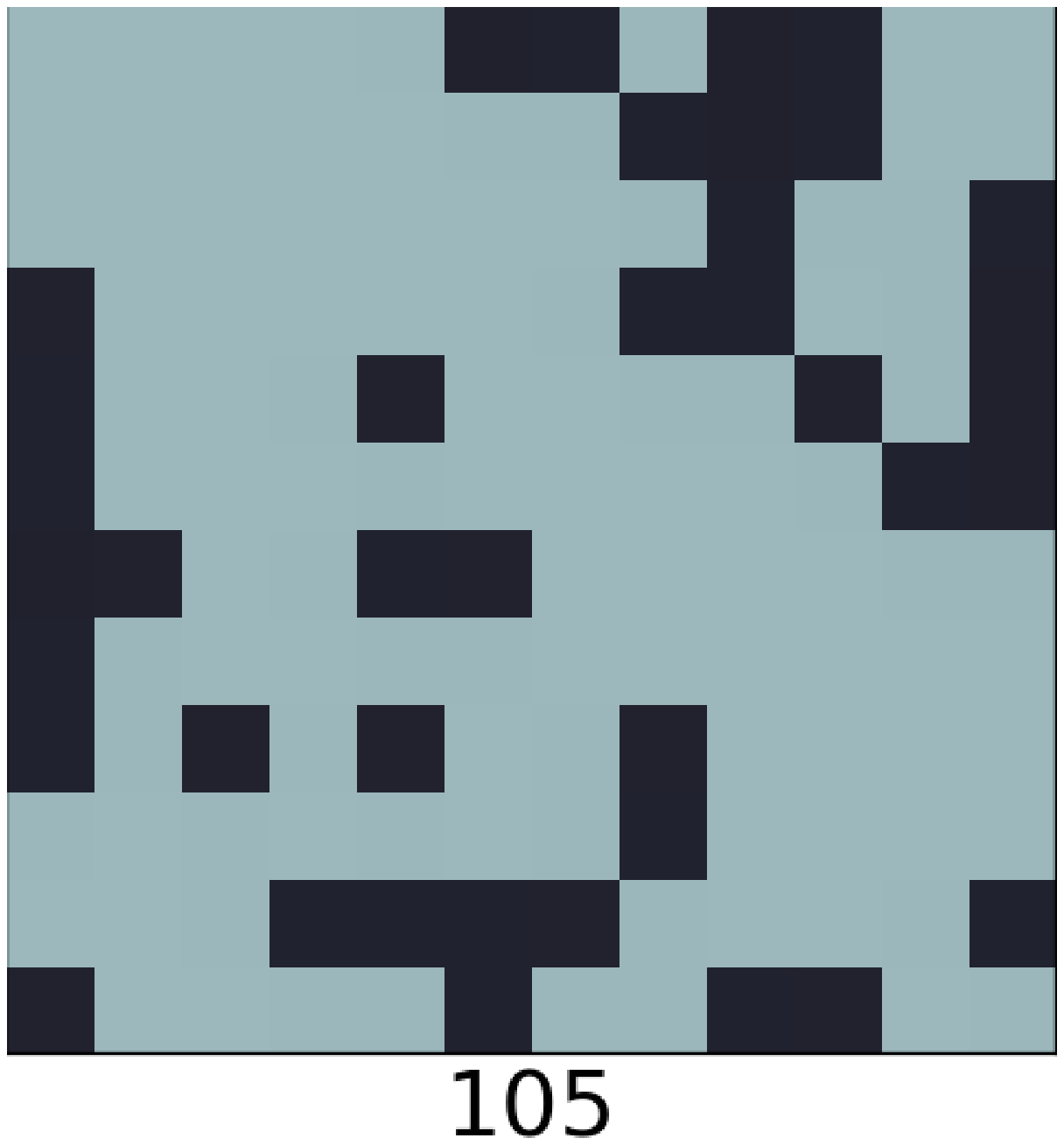}}\hspace{0.4em}
  \subfloat{\label{fig:snap38}\includegraphics[width=0.11\textwidth,height=0.8in]{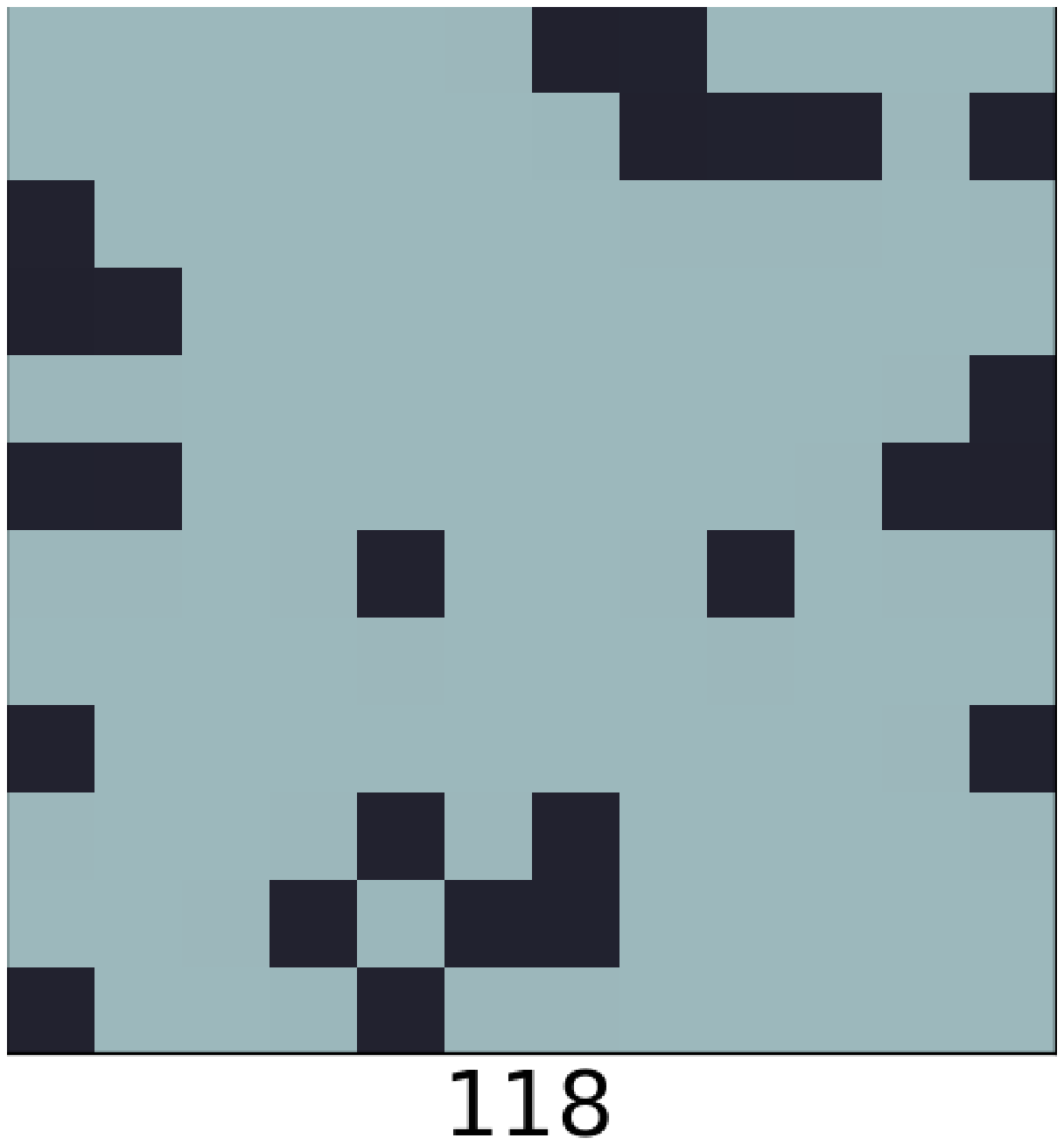}}\hspace{0.4em}
  \subfloat{\label{fig:snap39}\includegraphics[width=0.11\textwidth,height=0.8in]{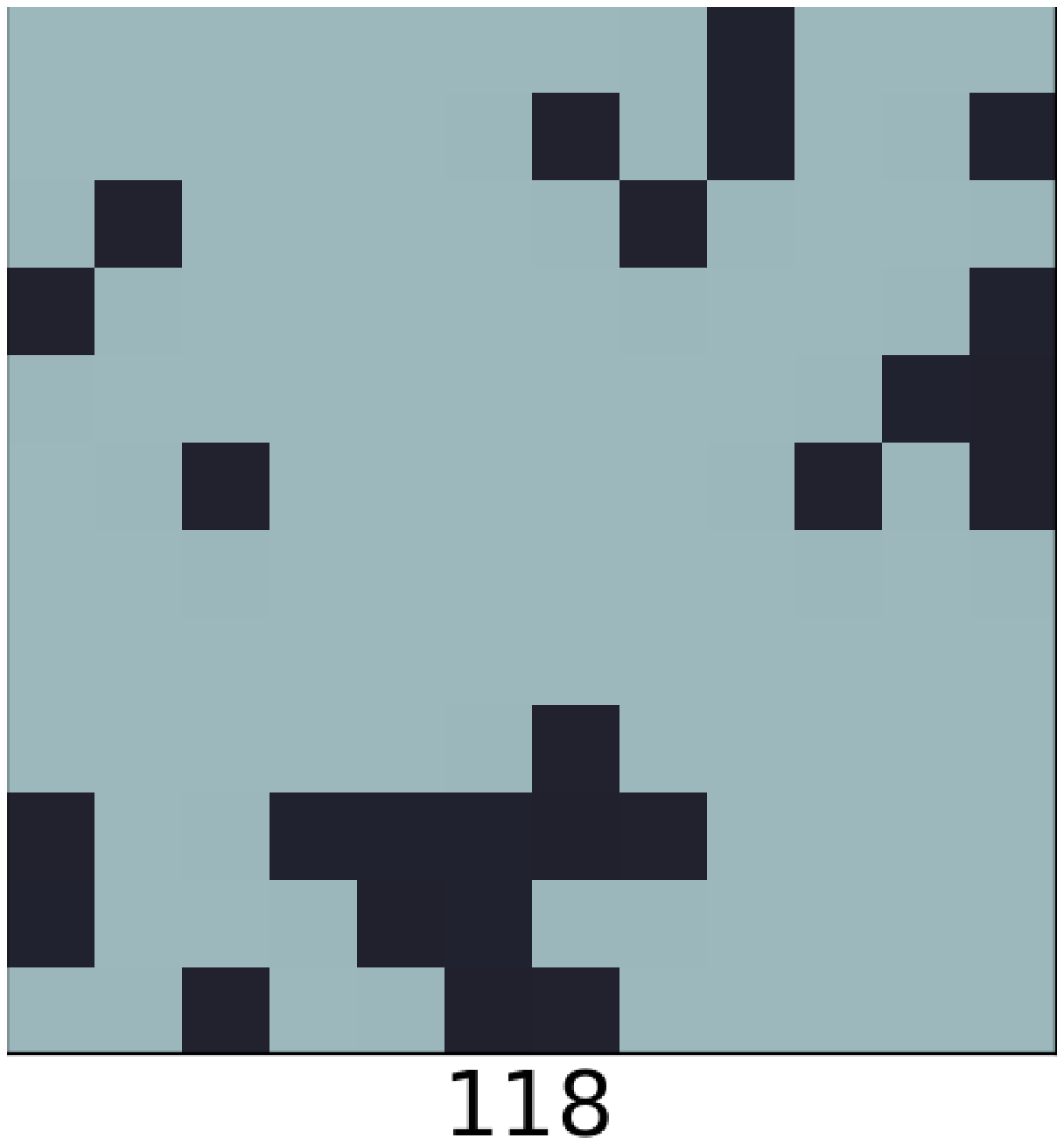}}\hspace{0.4em}\\

  \caption{Snapshots for first $40$ generations on $12\times 12$ grid and $c/b=0.45$. Defecting players are in dark. The labels give the size of the largest cooperative cluster.}
  \label{fig:snap}
\end{figure*}

While a single cooperative agent surrounded by defectors is always outperformed, cooperative players that are sufficiently 
inequity averse and adjacent to each other have two advantages: (1) the temporary clustering that arises due to sharing 
a common payoff and (2) the clustering provided by spatial constraints.
Like the previous model, in the initial games of a generation, if cooperative and defecting agents interact, the inequity in
accumulated payoff ensures that in the subsequent iterations the probability of C-D interactions decreases.
And as is traditionally the case, the spatial constraint
further improves the performance of cooperative players since players in the interior of the cluster enjoy the benefit of mutual cooperation.
Figure~\ref{fig:2dwithin_across} shows the fraction of interactions (for the  above sample run) within and outside the clusters 
for the first few generations. The corresponding snapshots for the first $40$ generations along with the size of the largest cooperative 
cluster (given below each plot) are shown in Figure.~\ref{fig:snap}. Defecting players are in black. We observe the cooperative players 
become dominant by generation $32$.

Figure~\ref{fig:2ddiffn} shows the fraction of cooperators vs $c/b$ for different grid sizes.
We observe $f_c$ does not change with $N$, and similar to the well-mixed model, cooperation disappears for $c/b \geq 0.5$.

\section{Conclusion}
We developed a model of agent interaction that is motivated by the social paradigm that individuals are inequity-averse and prefer to interact with others
within the same social strata. 
We presented  the results by considering both the well-mixed and the spatially structured populations across different parameter values. 
In general, cooperation becomes dominant when the cost of cooperation is low and is more robust for a structured population. 

Our results support the hypotheses that inequity aversion promotes cooperation among non kin \cite{evolve_ia}. 
It allows individuals to seek new partners with whom they can share more equitable payoffs. And if the equitable payoff increases the relative 
fitness of such  individuals (as is the case with the prisoner's dilemma), natural selection would guarantee that inequity averse cooperative agents emerge.
We also observe a strong correlation between inequity aversion and cooperation that points to the coevolution of these behaviors.
Brosnan \cite{evolve_ia} provides an extensive discussion of inequity aversion observed in other species \cite{range:2008,heidary:2008},
including capuchin monkeys \cite{brosnan:2003}, which ``elucidate evolutionary precursors to inequity aversion" \cite{brosnan:2004}.

We believe the model presented in this paper is an important step towards better understanding coevolution of cooperation 
and inequity aversion. In the future, we intend to evaluate the model by considering generic random networks and to incorporate other 
social factors like group membership, dominance rank, context of interaction, etc., 
which have been shown to effect the overall response to inequity \cite{evolve_ia}. 

\bibliographystyle{plain}
\bibliography{3reciprocal}

\begin{thebibliography}{10}

\bibitem{axelrod:1984}
Robert~M. Axelrod.
\newblock {\em The evolution of cooperation}.
\newblock Basic Books, New York, 1984.

\bibitem{boyd:1989}
Robert Boyd.
\newblock Mistakes allow evolutionary stability in the repeated prisoner's
  dilemma game.
\newblock {\em Journal of Theoretical Biology}, 136(1):47--56, Jan 1989.

\bibitem{brosnan:2003}
S.~F. Brosnan and F.~B.~M de~Waal.
\newblock Monkeys reject unequal pay.
\newblock {\em Nature}, 424:297--299, 2003.

\bibitem{evolve_ia}
Sarah~F Brosnan.
\newblock A hypothesis of the co-evolution of cooperation and responses to
  inequity.
\newblock {\em Frontiers in Neuroscience}, 5(43), 2011.

\bibitem{brosnan:2004}
Sarah~F. Brosnan and Frans B.~M de~Waal.
\newblock Animal behaviour: Fair refusal by capuchin monkeys.
\newblock {\em Nature}, 428(6979):140, 2004.

\bibitem{doebeli_2005}
Michael Doebeli and Christoph Hauert.
\newblock {Models of cooperation based on the Prisoner's Dilemma and the
  Snowdrift game}.
\newblock {\em Ecology Letters}, 8(7):748--766, July 2005.

\bibitem{dugatkin:1997}
L.A. Dugatkin.
\newblock {\em Cooperation among animals: an evolutionary perspective}.
\newblock Oxford Series in Ecology and Evolution Series. Oxford University
  Press, Incorporated, 1997.

\bibitem{Farrell_Ware:1989}
J.~Farrell and R.~Ware.
\newblock Evolutionary stability in the repeated prisoner's dilemma.
\newblock {\em Theoretical Population Biology}, 36(2):161--166, Oct 1989.

\bibitem{diffavers}
Ernest Fehr and Klaus~M Schmidt.
\newblock A theory of fairness, competition, and cooperation.
\newblock {\em Qarterly Journal of Economics}, 114(3):817--868, August 1999.

\bibitem{Fogel:1995}
David~B. Fogel.
\newblock On the relationship between the duration of an encounter and the
  evolution of cooperation in the iterated prisoner's dilemma.
\newblock {\em Evol. Comput.}, 3(3):349--363, September 1995.

\bibitem{fu_nowak2010}
Feng Fu, Martin~A. Nowak, and Christoph Hauert.
\newblock Invasion and expansion of cooperators in lattice populations:
  Prisoner's dilemma vs. snowdrift games.
\newblock {\em Journal of Theoretical Biology}, 266(3):358 -- 366, 2010.

\bibitem{goldberg:1991}
David~E. Goldberg and Kalyanmoy Deb.
\newblock A comparative analysis of selection schemes used in genetic
  algorithms.
\newblock In {\em Foundations of Genetic Algorithms}, pages 69--93. Morgan
  Kaufmann, 1991.

\bibitem{hamilton:1964}
W.~D. Hamilton.
\newblock {The genetical evolution of social behaviour. I}.
\newblock {\em Journal of Theoretical Biology}, 7(1):1--16, July 1964.

\bibitem{Hardin_1968_Commons}
Garrett Hardin.
\newblock The tragedy of the commons.
\newblock {\em Science}, 162:1243--1248, December 1968.

\bibitem{heidary:2008}
Fatemeh Heidary, Mohammad~Reza Vaeze~Mahdavi, Farshad Momeni, Bagher Minaii,
  Mehrdad Rogani, Nader Fallah, Roghayeh Heidary, and Reza Gharebaghi.
\newblock Food inequality negatively impacts cardiac health in rabbits.
\newblock {\em PLoS ONE}, 3(11):e3705, 11 2008.

\bibitem{langer:2008}
P~Langer, M~Nowak, and C~Hauert.
\newblock Spatial invasion of cooperation.
\newblock {\em J Theor Biol}, 250:634--641, 2008.

\bibitem{leimar:2001}
O.~Leimar and P.~Hammerstein.
\newblock {Evolution of cooperation through indirect reciprocity.}
\newblock {\em Proceedings. Biological sciences / The Royal Society},
  268(1468):745--753, April 2001.

\bibitem{maynardsmith:1973}
John {Maynard Smith} and George~R. Price.
\newblock {The Logic of Animal Conflict}.
\newblock {\em Nature}, 246(5427):15--18, November 1973.

\bibitem{NowakSigmund1993}
Martin Nowak and Karl Sigmund.
\newblock A strategy of win-stay, lose-shift that outperforms tit-for-tat in
  the prisoner's dilemma game.
\newblock {\em Nature}, 364(6432):56--58, 1993.

\bibitem{nowak_id:1998}
Martin Nowak and Karl Sigmund.
\newblock {Evolution of indirect reciprocity by image scoring}.
\newblock {\em Nature}, 393:573--7, June 1998.

\bibitem{nowakbook:2006}
Martin~A. Nowak.
\newblock {\em {Evolutionary Dynamics: exploring the Equations of Life}}.
\newblock Belknap Press of Harvard University Press, September 2006.

\bibitem{nowak_fiverules}
Martin~A. Nowak.
\newblock {Five Rules for the Evolution of Cooperation}.
\newblock {\em Science}, 314(5805):1560--1563, December 2006.

\bibitem{nowak_robert:1992}
Martin~A. Nowak and Robert~M. May.
\newblock {Evolutionary games and spatial chaos}.
\newblock {\em Nature}, 359(6398):826--829, October 1992.

\bibitem{nowak_nature:2005}
Martin~A. Nowak and Karl Sigmund.
\newblock Evolution of indirect reciprocity.
\newblock {\em Nature}, 437:1291--1298, 2005.

\bibitem{range:2008}
Friederike Range, Lisa Horn, Zs\'{o}fia Viranyi, and Ludwig Huber.
\newblock {The absence of reward induces inequity aversion in dogs.}
\newblock {\em Proceedings of the National Academy of Sciences of the United
  States of America}, 106(1):340--345, January 2009.

\bibitem{rc65}
A.~Rapoport and A.~M. Chammah.
\newblock {\em Prisoner's Dilemma: A Study in Conflict and Cooperation}.
\newblock University of Michigan Press, Ann Arbor, 1965.

\bibitem{riolo:1997}
Rick~L. Riolo.
\newblock The effects and evolution of tag-mediated selection of partners in
  populations playing the iterated prisoner's dilemma.
\newblock In {\em Seventh International Conference on Genetic Algorithms},
  pages 378--385, San Francisco, CA, USA, 1997. Morgan Kaufmann Publishers Inc.

\bibitem{riolo_axelrod:2001}
Rick~L. Riolo, Michael~D. Cohen, and Robert Axelrod.
\newblock {Evolution of cooperation without reciprocity}.
\newblock {\em Nature}, 414(6862):441--443, November 2001.

\bibitem{sigmund:2006}
K.~Sigmund.
\newblock {\em Games of life: explorations in ecology, evolution, and
  behaviour}.
\newblock Penguin science. Oxford University Press, 1993.

\bibitem{maynardsmith:1982}
John~Maynard Smith.
\newblock {\em Evolution and the Theory of Games}.
\newblock Cambridge University Press, Cambridge, UK, 1982.

\bibitem{suzuki:2005}
Shinsuke Suzuki and Eizo Akiyama.
\newblock {Reputation and the evolution of cooperation in sizable groups}.
\newblock {\em Proceedings of the Royal Society B: Biological Sciences},
  272(1570):1373--1377, July 2005.

\bibitem{szabo_tiffmode:1998}
Gy\"orgy Szab\'o and Csaba T\ifmmode~\mbox{\H{o}}\else \H{o}\fi{}ke.
\newblock Evolutionary prisoner's dilemma game on a square lattice.
\newblock {\em Phys. Rev. E}, 58:69--73, Jul 1998.

\bibitem{taylor:1992}
P.D. Taylor.
\newblock Altruism in viscous populations — an inclusive fitness model.
\newblock {\em Evolutionary Ecology}, 6(4):352--356, 1992.

\bibitem{traulsen_nowak:2006a}
A.~Traulsen and M.~A. Nowak.
\newblock {Evolution of cooperation by multilevel selection}.
\newblock {\em Proceedings of the National Academy of Sciences of the United
  States of America}, 103(29):10952--10955, July 2006.

\bibitem{trivers:1971}
Robert~L. Trivers.
\newblock {The Evolution of Reciprocal Altruism}.
\newblock {\em The Quarterly Review of Biology}, 46(1):35--57, 1971.

\bibitem{wilson:1975}
D.~S. Wilson.
\newblock {A theory of group selection.}
\newblock {\em Proceedings of the National Academy of Sciences of the United
  States of America}, 72(1):143--146, January 1975.

\end{thebibliography}
\end{document}